\documentclass[twocolumn]{autart}    

\usepackage{amsmath} 
\usepackage{amssymb}  
\usepackage{amsfonts}
\usepackage{mathtools}
\usepackage{color}
\usepackage{graphics} 
\usepackage{graphicx}
\usepackage{epsfig}   
\usepackage{epstopdf}
\usepackage[noadjust]{cite} 
\usepackage{fancyhdr}
\usepackage{fancyref}
\usepackage{hyperref}
\usepackage{float}
\usepackage{tikz}
\usetikzlibrary{calc,positioning,shapes,shadows,arrows,fit}
\usepackage{caption}
\usepackage{subcaption}
\usepackage{verbatim}

\newtheorem{problem}{Problem}

\newtheorem{lemma}{Lemma}
\newtheorem{proposition}{Proposition}
\newtheorem{remark}{Remark}

\newtheorem{assumption}{Assumption}

\newcommand{\scr}{\scriptscriptstyle}

\usepackage[font = footnotesize, skip=3pt]{caption}

\begin{document}

\begin{frontmatter}

\title{Robust Formation Control in $\mathbb{SE}(3)$ for Tree-Graph Structures with Prescribed Transient and Steady State Performance} 

\thanks[footnoteinfo]{This work was supported by the H2020 ERC Starting Grant BUCOPHSYS, the European Union's Horizon 2020 Research and Innovation Programme under the Grant Agreement No. 644128 (AEROWORKS), the EU H2020 Research and Innovation Programme under GA No. 731869 (Co4Robots), the Swedish Research Council (VR), the Knut och Alice Wallenberg Foundation (KAW) and the Swedish Foundation for Strategic Research (SSF). \emph{Email addresses:} \{cverginis, anikou, dimos\}@kth.se }

\author{Christos K. Verginis},    
\author{Alexandros Nikou}, 
\author{Dimos V. Dimarogonas}

\address{KTH Center of Autonomous Systems and ACCESS Linnaeus Center, School of Electrical Engineering and Computer Science, KTH Royal Institute of Technology, SE-100 44, Stockholm, Sweden.}  

                                                                      

\begin{abstract}
This paper presents a novel control protocol for distance and orientation formation control of rigid bodies, whose sensing graph is a static and undirected tree, in the special Euclidean group $\mathbb{SE}(3)$. The proposed control laws are decentralized, in the sense that each agent uses only local relative information from its neighbors to calculate its control signal, as well as robust with respect to modeling (parametric and structural) uncertainties and external disturbances. The proposed methodology guarantees the satisfaction of inter-agent distance constraints that resemble collision avoidance and connectivity maintenance properties. Moreover, certain predefined functions characterize the transient and steady state performance of the closed loop system. Finally, simulation results verify the validity and efficiency of the proposed approach.

\end{abstract}

\end{frontmatter}

\section{Introduction}

During the last decades, decentralized control of multi-agent systems has gained a significant amount  of  attention  due  to  the  great  variety  of  its  applications,  including  multi-robot  systems, transportation,  multi-point  surveillance  as well as  biological  systems. Among the various research topics in multi-agent systems, the most popular ones can be considered to be (i) multi-agent navigation \cite{dimarogonas2007decentralized}, where the agents need to navigate to predefined positions of the state space, and (ii) consensus \cite{olfati_murray_concensus}, where the agents aim to converge to a common state. At the same time, the agents might need to fulfill certain transient properties, such as network connectivity \cite{magnus_2007_connectivity} and/or collision avoidance \cite{dimos_2006_automatica_nf}. 
Another important problem considered in multi-agent systems is formation control \cite{oh_park_ahn_2015}, where the agents aim to form a predefined shape in the state space, and which can be seen as a combination of the navigation and consensus problems. Formation control is categorized  in (\cite{oh_park_ahn_2015}) position-based, distance-based and orientation-based formation control, as well as a combination of the two, which is also the focus of this work.


Distance-based formation control has been well-studied in the related literature (see, indicatively, \cite{anderson_yu_fidan_hendrickx_2008, yu_anderson_dagsputa_fidan_2009, krick_broucke_francis_2009, dorfler_francis_2010, oh_ahn_2011e, cao_morse_yu_anderson_dagsputa_2011, summers_yu_dagsputa_anderson_2011_tac,  belabbas2012robustness, zelazo_2015, Ferreira-Vazquez2016, bdo_2016_distance_based_formation,dimarogonas2010stability}). In these works, however, the authors consider simplified single-integrator models for the agent dynamics. Double integrator schemes have been studied in \cite{magnus_formation, olfati_murray_2002, oh_ahn_2014a}. Orientation-based formation control has been investigated in \cite{basiri_2010_angle_formation, eren_2012_bearing_formation, zhao2016bearing, oh_ahn_2014_angle_based_formation}, whereas the authors in \cite{oh_ahn_2014_angle_based_formation, bishop_2015_distributed, fathian2016globally} have considered the combination of distance- and orientation-based formation, also employing single integrator or $2$D unicycle dynamics.

The use of simplified dynamics however, like in the aforementioned works, does not apply to realistic engineering applications, where the systems may have complicated and uncertain dynamics. Moreover, such systems are inherently under the presence of exogenous disturbances. Two more characteristics not taken into account in most of the aforementioned works is (i) connectivity preservation among the agents, and (ii) inter-agent collision avoidance. Both of these properties are important, inherent from the limited sensing capabilities of
multi-agent systems, and dimensionless agents/robots in potential real-time applications, respectively.  

Motivated by the above, we present in this paper a novel control protocol for the formation control of multiple rigid bodies forming a tree sensing graph in $\mathbb{SE}(3)$. We employ the Prescribed Performance Control methodology, initially proposed in \cite{bechlioulis_tac_2008}, to achieve predefined transient- and steady-state performance. Prescribed performance control has been considered in the framework of multi-agent systems in \cite{babis_2014_formationCDC,babis_2014_formationICRA,bechlioulis2017_masTAC,dimos_ppc_consensus}. In \cite{babis_2014_formationCDC,babis_2014_formationICRA} the authors tackle the position-based formation control problem, by taking into-account position-based connectivity maintenance in \cite{babis_2014_formationICRA}, and \cite{bechlioulis2017_masTAC,dimos_ppc_consensus} consider the consensus problem. The proposed methodology exhibits the following attributes: $1)$ It is decentralized, in the sense that each agent computes its own control signal based on its local sensing capabilities, without needing to communicate with the rest of the agents, or to know the pose of a global coordinate frame. $2)$ It is robust to bounded external disturbances and uncertainties of the dynamic model, since these are not employed in the control design. $3)$ It guarantees satisfaction of certain distance constraints among the initially connected agents, which resemble collision avoidance and connectivity maintenance specifications. $4)$ It guarantees convergence to a feasible formation configuration with predefined transient and steady-state performance from almost all initial conditions. Moreover, in contrast to standard continuous control methodologies on $\mathbb{SO}(3)$ (where the closer the initial condition is to the unstable equilibrium, the more the stabilization time approaches infinity), it guarantees convergence to the formation configuration \textit{arbitrarily fast}, regardless of the distance of the initial system configuration to the unstable equilibrium. This paper constitutes an extension of our previous works \cite{alex_chris_ppc_formation_ifac}, \cite{chris_alex_cdc}. In both of these works  we addressed the same problem using Euler angles that suffer from representation singularities as well as knowledge of a common global inertial frame; \cite{chris_alex_cdc} employs a potential function-based solution, inherently exhibiting local minima, and  \cite{alex_chris_ppc_formation_ifac} also uses the idea of prescribed performance control.

%


\section{Notation and Preliminaries} \label{sec:preliminaries}

The set of positive integers is denoted as $\mathbb{N}$. The real $n$-coordinate space, with $n\in\mathbb{N}$, is denoted as $\mathbb{R}^n$; $\mathbb{R}^n_{\geq 0}$ and $\mathbb{R}^n_{> 0}$ are the sets of real $n$-vectors with all elements nonnegative and positive, respectively. Given a set $S$, denote by $\lvert S\lvert$ its cardinality, by $S^n = S \times \dots S$ its $n$-fold Cartesian product, and by $2^S$ the set of all its subsets. The notation $\|x\|$ is used for the Euclidean norm of a vector $x \in \mathbb{R}^n$. Given a symmetric matrix $A, \lambda_{\text{min}}(A) \coloneqq \min \{|\lambda| : \lambda \in \text{eig}(A) \}$ denotes the minimum eigenvalue of $A$, respectively, where $\text{eig}(A)$ is the set of all the eigenvalues of $A$ and $\text{rank}(A)$ is its rank;  $\|A\|_\text{F} \coloneqq \text{tr}(A^\top A)$ is the Frobenius norm of $A$, and $\text{tr}[\cdot]$ is its trace; \text{det}(A) denotes the determinant of a matrix $A \in \mathbb{R}^{n \times n}$. The notation $\text{diag}\{A_1, \dots, A_n\}$ stands for the block diagonal matrix with the matrices $A_1$, $\dots$, $A_n$ in the main block diagonal; $A \otimes B$ denotes the Kronecker product of matrices $A, B \in \mathbb{R}^{m \times n}$, as was introduced in \cite{horn_jonshon}. Define by $I_n \in \mathbb{R}^{n \times n}$ and $0_{m \times n} \in \mathbb{R}^{m \times n}$ the unitary matrix and the $m \times n$ matrix with all entries zeros, respectively; $\mathcal{B}(c,r) \coloneqq \{x \in \mathbb{R}^3: \|x-c\| \leq r\}$ is the vector-valued mapping representing the $3$D ball of radius $r\in\mathbb{R}_{>0}$ and center $c\in\mathbb{R}^{3}$.  Given $x$, $y\in\mathbb{R}^3$, $S:\mathbb{R}^{3} \to  \mathfrak{so}(3)$ is the skew-symmetric matrix defined according to $S(x)y = x\times y$, and $S^{-1}: \mathfrak{so}(3)\to\mathbb{R}^3$ is its inverse, where $\mathfrak{so}(3) = \{S\in\mathbb{R}^{3\times 3} : x^\top S(\cdot) x = 0, \forall x \in \mathbb{R}^3\}$ is the space of skew-symmetric matrices. The special Euclidean group is denoted by $\mathbb{SE}(3) \coloneqq \{ (c,R) \in \mathbb{R}^3\times \mathbb{SO}(3) \}$, where $\mathbb{SO}(3) \coloneqq \{R\in\mathbb{R}^{3\times 3} : R^\top R = I_3, \text{det}(R) = 1\}$. Moreover, the tangent space to $\mathbb{SO}(3)$ at $R$ is denoted by $T_R\mathbb{SO}(3)$ and we also use $\mathbb{T}_{R} \coloneqq \mathbb{R}^3\times T_R\mathbb{SO}(3)$. We define the induced norm in $\mathbb{SO}(3)^N$ as $\| R \|_{T} \coloneqq \sum_{i\in\{1,\dots,N\}}\|R_i\|_{\text{F}}$ for any $R = (R_1,\dots,R_N)\in\mathbb{SO}(3)^N$.
Finally, all the differentiations are performed with respect to an inertial frame of reference unless otherwise stated. Some useful properties of skew symmetric matrices \cite{lee2010control}: $x^\top S(y) x = 0;	
S(R x) = R S(x) R^\top,	
-\frac{1}{2} \text{tr} \left[ S(x) S(y) \right] = x^\top y,
\text{tr} \left[A S(x) \right] = \frac{1}{2} \text{tr} \left[ S(x) (A - A^\top) \right] = - x^\top S^{-1}(A-A^\top)$, 
for every $x$, $y \in \mathbb{R}^{3}$, $A \in \mathbb{R}^{3 \times 3}$ and $R \in \mathbb{SO}(3)$.

\subsection{Prescribed Performance Control}

\label{subsec:ppc}
Prescribed Performance Control (PPC), originally proposed in \cite{bechlioulis_tac_2008}, describes the behavior where a tracking error $e(t):\mathbb{R}_{\geq 0}  \to  \mathbb{R}$ evolves strictly within a predefined region that is bounded by certain functions of time, achieving prescribed transient and steady state performance.
The mathematical expression of prescribed performance is given by the following inequalities: $-\rho_{\scriptscriptstyle L}(t) < e(t) < \rho_{\scriptscriptstyle U}(t),\ \ \forall t\in\mathbb{R}_{\geq 0}, \notag$ where $\rho_{\scriptscriptstyle L}(t),\rho_{\scriptscriptstyle U}(t)$ are smooth and bounded decaying functions of time, satisfying $\lim\limits_{t  \to  \infty}\rho_{\scriptscriptstyle L}(t) > 0$ and $\lim\limits_{t  \to  \infty}\rho_{\scriptscriptstyle U}(t) > 0$, called performance functions. Specifically, for the exponential performance functions $\rho_i(t) = (\rho_{i 0}-\rho_{i \infty})e^{-l_it}+\rho_{i \infty}$, with $\rho_{i 0},\rho_{i \infty}, l_i\in\mathbb{R}_{>0}, i\in\{U,L\}$, appropriately chosen constants, $\rho_{\scriptscriptstyle L 0}=\rho_{\scriptscriptstyle L}(0)$, $\rho_{\scriptscriptstyle U 0}=\rho_{\scriptscriptstyle U}(0)$ are selected such that $\rho_{\scriptscriptstyle U 0} > e(0) > \rho_{\scriptscriptstyle L 0}$ and the constants $\rho_{\scriptscriptstyle L \infty}=\lim\limits_{t  \to  \infty} \rho_{\scriptscriptstyle L}(t)< \rho_{\scriptscriptstyle L 0}$, $\rho_{\scriptscriptstyle U \infty}=\lim\limits_{t  \to  \infty}\rho_{\scriptscriptstyle U}(t)<\rho_{\scriptscriptstyle U 0}$ represent the maximum allowable size of the tracking error $e(t)$ at steady state, which may be set arbitrarily small to a value reflecting the resolution of the measurement device, thus achieving practical convergence of $e(t)$ to zero. Moreover, the decreasing rate of $\rho_{\scriptscriptstyle L}(t)$, $\rho_{\scriptscriptstyle U}(t)$, which is affected by the constants $l_{\scriptscriptstyle L}$, $l_{\scriptscriptstyle U}$ in this case, introduces a lower bound on the required speed of convergence of $e(t)$. Therefore, the appropriate selection of the performance functions $\rho_{\scriptscriptstyle L}(t)$, $\rho_{\scriptscriptstyle U}(t)$ imposes performance characteristics on the tracking error $e(t)$.


\subsection{Dynamical Systems} \label{subsec:dynamical systems}

\begin{thm} \label{thm:ode solution} \cite[Theorem 2.1.1]{bressan2007introduction}
Let $\Omega$ be an open set in $\mathbb{R}^n\times\mathbb{R}_{\geq 0}$. Consider a function $g:\Omega  \to  \mathbb{R}^n$ that satisfies the following conditions: $1)$ For every $z\in\mathbb{R}^n$, the function $t \to  g(z,t)$ defined on $\Omega_z \coloneqq \{t : (z,t)\in\Omega\}$ is measurable. For every $t\in\mathbb{R}_{\geq 0}$, the function $z \to  g(z,t)$ defined on $\Omega_t \coloneqq \{z : (z,t)\in\Omega\}$ is continuous; $2)$ For every compact $S\subset \Omega$, there exist constants $C_S$, $L_S$ such that: $\|g(z,t) \| \leq C_S, \|g(z,t)-g(y,t) \| \leq L_S \|z-y \|$, $\forall (z,t),(y,t) \in S$. Then, the initial value problem $\dot{z} = g(z,t)$, $z_0=z(t_0)$, for some $(z_0,t_0)\in\Omega$, has a unique and maximal solution defined in $[t_0,t_{\max})$, with $t_{\max} > t_0$ such that $(z(t),t)\in\Omega, \forall t\in[t_0,t_{\max})$.
\end{thm}

\begin{thm} \label{thm:forward_completeness} \cite[Theorem 2.1.4]{bressan2007introduction}
Let the conditions of Theorem \ref{thm:ode solution} hold in $\Omega$ and let a maximal solution of the initial value problem $\dot{z} = g(z,t)$, $z_0=z(t_0)$, exists in $[t_0,t_{\max})$ such that $(z(t),t)\in\Omega, \forall t\in[t_0,t_{\max})$. Then, either $t_{\max} = \infty$ or $\lim\limits_{t \to  t^{-}_{\max}}\Big[\| z(t) \| + \frac{1}{d_\mathcal{S}((z(t),t),\partial \Omega)} \Big] = \infty$,
where $d_{\mathcal{S}}: \mathbb{R}^n\times2^{\mathbb{R}^n}\to\mathbb{R}_{\geq 0}$ is the distance of a point $x\in\mathbb{R}^n$ to a set $A$, defined as $d_{\mathcal{S}}(x,A) \coloneqq \inf\limits_{y\in A}\{\|x-y\|\}$.
\end{thm}

\subsection{Graph Theory} \label{subsection: graph theory}

An \textit{undirected graph} $\mathcal{G}$ is a pair $(\mathcal{N}, \mathcal{E})$, where $\mathcal{N}$ is a finite set of $N\in\mathbb{N}$ nodes, representing a team of agents, and $\mathcal{E} \subseteq \{ \{i,j\} : \forall i,j \in \mathcal{N}, i \neq j\}$, with $K \coloneqq |\mathcal{E}|$, is the set of edges that model the sensing capabilities between neighboring agents. For each agent, its neighboring set $\mathcal{N}_i$ is defined as $\mathcal{N}_i \coloneqq \{ j \in \mathcal{N}: \{i,j\} \in \mathcal{E}\}$.
If there is an edge $\{i, j\} \in \mathcal{E}$, then $i, j$ are called \textit{adjacent}. A \textit{path} of length $r$ from vertex $i$ to vertex $j$ is a sequence of $r+1$ distinct vertices, starting with $i$ and ending with $j$, such that consecutive vertices are adjacent. For $i = j$, the path is called a \text{cycle}. If there is a path between any two vertices of the graph $\mathcal{G}$, then $\mathcal{G}$ is called \textit{connected}. A connected graph is called a \text{tree} if it contains no cycles.
Consider an arbitrary orientation of $\mathcal{G}$, which assigns to each edge $\{i, j\} \in \mathcal{E}$ precisely one of the ordered pairs $(i, j)$ or $(j, i)$. When selecting the pair $(i, j)$, we say that $i$ is the tail and $j$ is the head of the edge $\{i, j\}$. By considering a numbering $k \in \mathcal{K} \coloneqq \{1, \dots , K\}$ of the graph's edge set, we define the $N\times K$ \textit{incidence matrix} $D(G) = [d_{ij}]$, where: $d_{ij} = 1$, if $i$ is the head of edge $j$; $d_{ij} = -1$, if $i$ is the tail of edge $j$; and $d_{ij} = 0$, otherwise. 

\begin{lemma} \label{lemma:tree} \cite[Section III]{dimarogonas2010stability}
	Assume that the graph $\mathcal{G}$ is a tree. Then, $D(\mathcal{G})^\top \Delta D(\mathcal{G})$ is positive definite for any positive definite matrix $\Delta\in\mathbb{R}^{N\times N}$.
\end{lemma}

\begin{proposition} \label{prop f(x)}
	Let $f:\mathbb{R}_{\geq 0} \to  \mathbb{R}$, with $f(x) \coloneqq \exp(x)\left[\exp(x)-1\right] - x^2$. Then it holds that $f(x) \geq 0$, $\forall x\in\mathbb{R}_{\geq 0}$.
\end{proposition}

\begin{proposition} \label{prop:e_R frobenious} \cite{lee_2017}
	Let $R_1, R_2 \in \mathbb{SO}(3)$, and $e_R \coloneqq S^{-1}( R^\top_1 R_2 - R^\top_2 R_1)$. Then $\|e_R \|^2  \coloneqq \|R_1 - R_2 \|^2_\text{F}\Big(1 - \tfrac{1}{8}\|R_1 - R_2 \|^2_\text{F} \Big)$.
\end{proposition}

\begin{proposition} \cite{lee_attitude_control_letters} \label{prop:R trace}
	Let $R_1, R_2 \in \mathbb{SO}(3)$. Then, for the rotation matrix $R_2^\top R_1 \in \mathbb{SO}(3)$ it holds that $-1 \le \text{tr}[R_2^\top R_1] \le 3$; $\text{tr}[R_2^\top R_1] = 3$ if and only if $R_2^\top R_1= I_3 \Leftrightarrow R_1 = R_2$; $\text{tr}[R_2^\top R_1] = -1$ when $R_1 = R_2 \exp(\pm\pi S(x))$, for every $x$ in the unit sphere, where $\exp(\cdot)$ here is the matrix exponential.
\end{proposition}

\section{Problem Formulation} \label{sec:prob_formulation}

Consider a set of $N$ rigid bodies, with $\mathcal{N} = \{ 1,2, \ldots, N\}$, $N  \geq 2$, operating in a workspace $W\subseteq \mathbb{R}^3$. We consider that each agent occupies a ball $\mathcal{B}(p_i,r_i)$, where $p_i\in\mathbb{R}^3$ is the position of the agent's center of mass with respect to an inertial frame $\mathcal{F}_o$ and $r_i\in\mathbb{R}_{>0}$ is the agent's radius. 
We also denote as $R_i\in \mathbb{SO}(3)$ the rotation matrix associated with the orientation of the $i$th rigid body.  Moreover, we denote by $v_{i,L}\in\mathbb{R}^3$ and $\omega_i\in\mathbb{R}^3$ the linear and angular velocity of agent $i$ with respect to frame $\mathcal{F}_o$. The vectors $p_i$ are expressed in $\mathcal{F}_o$ coordinates, whereas $v_{i,L}$ and $\omega_i$ are expressed with respect to a local frame $\mathcal{F}_i$ centered at each agent's center of mass. The position of $\mathcal{F}_o$, though, is not required to be known by the agents, as will be shown later.
By defining $x_i \coloneqq (p_i, R_i) \in \mathbb{SE}(3)$ and $v_i\coloneqq [v_{i,L}^\top, \omega^\top_i]^\top\in\mathbb{R}^6$, we model each agent's motion with the $2$nd order Newton-Euler dynamics:
\begin{subequations}\label{eq:system} 
	\begin{align} 
	&\hspace{-3mm} \dot{x}_i = (R_iv_{i,L}, R_i S(\omega_i)) \in \mathbb{T}_{R_i}, \label{eq:system_1} \\ 
	&\hspace{-3mm} u_i = M_i \dot{v}_i + C_i(v_i)v_i+g_i(x_i) + w_i(x_i,v_i,t),  \label{eq:system_2} 
	\end{align}
\end{subequations}
where the matrix $M_i\in \mathbb{R}^{6\times6}$ is the constant positive definite inertia matrix, $C_i: \mathbb{R}^6 \to  \mathbb{R}^{6\times6}$ is the Coriolis matrix, $g_i:\mathbb{SE}(3)\to\mathbb{R}^6$ is the body-frame gravity vector,  $w_i:\mathbb{SE}(3)\times\mathbb{R}^6\times\mathbb{R}_{\geq 0}  \to  \mathbb{R}^6$ is a bounded vector representing model uncertainties and external disturbances, and $\mathbb{T}_{R_i} = \mathbb{R}^3\times T_{R}\mathbb{SO}(3)$, as defined in Section \ref{sec:preliminaries}. Finally, $u_i\in\mathbb{R}^6$ is the control input vector representing the $6$D body-frame generalized force acting on agent $i$. The following properties hold for the aforementioned terms:
\begin{itemize}
	\item The terms $M_i, C_i(\cdot), g_i(\cdot)$ are \textit{unknown} to the agents, $C_i(\cdot), g_i(\cdot)$ are continuous, and it holds that 
	\begin{subequations} \label{eq:M property}
		
	\begin{align}
		 0 & < \underline{m}_i < \bar{m}_i < \infty \label{eq:M property 1}\\
			\|g_i(x_i)\| & \leq \bar{g}_i, \forall x_i\in\mathbb{SE}(3), \label{eq:M property 2}
	\end{align}
	\end{subequations}
	where $\bar{g}_i$ is a finite \textit{unknown} positive constant and $\underline{m}_i\coloneqq \lambda_{\min}(M_i)$, and $\bar{m}_i \coloneqq \lambda_{\max}(M_i)$, which are also \textit{unknown} to the agents, $\forall i\in\mathcal{N}$. 
	\item  The functions $w_i(x_i,v_i,t)$ are assumed to be continuous in $v_i\in\mathbb{R}^6$ and bounded in $(x_i,t)$ by \textit{unknown} positive finite constants $\bar{w}_i$.
\end{itemize}

The dynamics \eqref{eq:system_2} can be written in a vector form representation as:
\begin{subequations} \label{eq:system_MAS}
\begin{align} 
\dot{x} & = h_x(x,v),\\ 
u & = M \dot{v} + C(v) v +g(x) + w(x,v,t), 
\end{align}
\end{subequations}
where $x \coloneqq (x_1,\dots,x_N)\in \mathbb{SE}(3)^N$, $v \coloneqq [v_1^\top, \dots, v_N^\top]^\top$ $\in \mathbb{R}^{6N}$, $u \coloneqq [u_1^\top, \dots, u_N^\top]^\top \in \mathbb{R}^{6N}$, and $h_{x}(x,v)$ $\coloneqq$ $(h_{x_1}(x_1,v_1)$, $\dots$, $h_{x_N}(x_N,v_N))$
$\coloneqq$ $((R_1v_{1,L}$, $R_1S(\omega_1))$, $\dots$, $(R_Nv_{N,L}$, $R_NS(\omega_N)) )$ 
$\in$ $\mathbb{T}_{R_1}\times\dots\times\mathbb{T}_{R_N}$, $M$ $\coloneqq$ $\text{diag}\{[M_i]_{i\in\mathcal{V}} \} \in\mathbb{R}^{6N\times 6N}$, $C(v)$ $\coloneqq$ $\text{diag}\{[C_i(v_i)]_{i\in\mathcal{V}}\}$ $\in\mathbb{R}^{6N\times 6N}$, 
 $g(x)$ $\coloneqq$ $[g_1(x_1)^\top,\dots, g_N(x_N)^\top]^\top$ $\in\mathbb{R}^{6N}$,  
 $w(x,v,t)$ $\coloneqq$ $[w_1(x_1,v_1,t)^\top,\dots, w(x_N,v_N,t)^\top]^\top$ $\in\mathbb{R}^{6N}$.

It is also further assumed that each agent has a limited sensing range of $s_i > \max_{i,j\in\mathcal{N}}\{r_i+r_j\}$. Therefore, by defining the neighboring function 
$\mathcal{N}_i(p) \coloneqq \{j\in\mathcal{N} : p_j\in\mathcal{B}(p_i,s_i)\}$, and $p\coloneqq [p^\top_1,\dots,p^\top_N]^\top\in\mathbb{R}^{3N}$, agent $i$ can measure the relative offset $R^\top_i(p_i - p_j)$ (i.e., expressed in $i$'s local frame), the distance $\|p_i - p_j\|$, as well as the relative orientation $R^\top_j R_i$ with respect to its neighbors $j\in\mathcal{N}_i(p)$. In addition, we consider that each agent can measure its own velocity subject to time- and state-varying bounded noise, i.e., agent $i$ has continuous feedback of $\widetilde{v}_i \coloneqq [\widetilde{v}_{i,L}^\top,\widetilde{\omega_i}^\top]^\top \coloneqq v_i + n_i(x_i,t)$, $\forall i\in\mathcal{N}$; $n_i(x_i,t)$ are assumed to be bounded by \textit{unknown} positive finite constants $\bar{n}_i$ and $n_{i,d}(x_i,\dot{x}_i,t) \coloneqq\dot{n}_i(x_i,t)$ are assumed to be continuous in $\dot{x}_i$ and bounded in $(x_i,t)$ by \textit{unknown} positive finite constants $\bar{n}_{i,d}$, $\forall i\in\mathcal{N}$.
	

\begin{remark} $[$\textbf{Local relative feedback}$]$
		Note that the agents do not need to have information of any common global inertial frame. The feedback they obtain is relative with respect to their neighboring agents (expressed in their local frames) and they are not required to perform transformations in order to obtain absolute positions/orientations. In the same vein, note also that the velocities $v_i$ are vectors expressed in the agents' local frames.
\end{remark}

The topology of the multi-agent network is modeled through the \textit{undirected} graph $\mathcal{G} = (\mathcal{N},\mathcal{E})$, with $\mathcal{E}=\{(i,j)\in\mathcal{N}^2 : j\in\mathcal{N}_i(p(0)) \text{ and } i\in\mathcal{N}_j(p(0))\}$ (i.e., the set of initially connected agents), which is assumed to be nonempty and \textit{connected}. We further denote $\mathcal{K} \coloneqq \{1,\dots,K\}$ where $K \coloneqq |\mathcal{E}|$. Given the $k$-th edge, we use the simplified notation $(k_1,k_2)$ for the function that assigns to edge $k$ the respective agents, with $k_1,k_2\in\mathcal{N}$, $\forall k\in\mathcal{K}$. 
Since the agents are heterogeneous with respect to their sensing capabilities (different sensing radii $s_{i}$), the fact that the initial graph is nonempty, connected and undirected implies that
\begin{equation} \label{eq:connect at t=0}
\lVert p_{k_2}(0)-p_{k_1}(0) \rVert < d_{k,\text{con}},
\end{equation}
with $d_{k,\text{con}} \coloneqq \min\{s_{k_1},s_{k_2}\},\forall k\in \mathcal{K}$. In other words, we consider that the position of the agents at $t=0$ is such that the agents for which \eqref{eq:connect at t=0} holds form a connected sensing graph.
We also consider that $\mathcal{G}$ is static in the sense that no edges are added to the graph. We do not exclude, however, edge removal through connectivity loss between initially neighboring agents, which we guarantee to avoid. That is, the proposed methodology guarantees that $\lVert p_{k_2}(t)-p_{k_1}(t) \rVert < d_{k,\text{con}}$, $\forall k\in\mathcal{K}$, $\forall t\in\mathbb{R}_{\geq 0}$.
It is also assumed that at $t=0$ the neighboring agents are at a collision-free configuration, i.e., $d_{k,\text{col}} < \lVert p_{k_2}(0)-p_{k_1}(0)\rVert, \forall k\in \mathcal{K}$, with $d_{k,\text{col}} \coloneqq r_{k_1}+r_{k_2}$. Hence, we conclude that 
\begin{equation}
d_{k,\text{col}} < \lVert p_{k_2}(0)-p_{k_1}(0)\rVert < d_{k,\text{con}}, \forall k\in \mathcal{K}. \label{eq:at t=0}
\end{equation}
The desired formation is specified by the constants $d_{k,\text{des}}\in\mathbb{R}_{\geq 0}, R_{k,\text{des}}\in\mathbb{SO}(3), \forall k\in \mathcal{K}$, for which, the formation configuration is called \textit{feasible} if the set ${\Phi} \coloneqq \{x\in \mathbb{SE}(3)^N  : \lVert p_{k_2} - p_{k_1} \rVert = d_{k,\text{des}}, \ R^\top_{k_2}R_{k_1} = R_{k,\text{des}}, \forall k\in \mathcal{K}\}$ is nonempty.   
Apart from achieving a desired inter-agent formation while maintaining the initial edges, we aim at guaranteeing that the inter-agent distance of the edges $k\in\mathcal{K}$ (initially connected agents) stays larger than $r_{k_1} + r_{k_2}$, complying with potential collision avoidance specifications.
We also make the following required assumption: 
\begin{assumption} \label{assump:basic_assumption}
	The sensing graph $\mathcal{G}$ is a tree.
\end{assumption}
The aforementioned assumption states the initially connected agents in $\mathcal{E}$ must form a tree graph. In cases where the agents satisfying \eqref{eq:connect at t=0} form a graph that contains cycles, edges can be manually deleted according to certain criteria (e.g. neighboring priorities) in order to obtain a tree sensing graph.

\begin{problem} \label{problem}
Given $N$ agents governed by the dynamics \eqref{eq:system}, under Assumption \ref{assump:basic_assumption} and given the desired inter-agent configuration constants $d_{k,\text{des}}\in\mathbb{R}_{\geq 0}$, $R_{k,\text{des}}\in\mathbb{SO}(3)$, with $d_{k,\text{col}}<d_{k,\text{des}} < d_{k,\text{con}}$, $\forall k\in \mathcal{K}$, design decentralized control laws $u_i \in\mathbb{R}^6, i\in\mathcal{N}$ such that, $\forall \ k \in \mathcal{K}$, the following hold: $1)$ $\lim\limits_{t  \to  \infty} \|p_{k_2}(t)-p_{k_1}(t)\| = d_{k,\text{des}}$; $2)$ $\lim\limits_{t  \to  \infty} [R_{k_2}(t)]^\top R_{k_1}(t) = R_{k,\text{des}}$; $3)$ $d_{k,\text{col}} < \|p_{k_2}(t)-p_{k_1}(t)\| < d_{k,\text{con}}, \forall \ t \in \mathbb{R}_{\geq 0}$.
\end{problem}

The term ``robust" here refers to robustness of the proposed methodology with respect to the unknown dynamics and external disturbances in \eqref{eq:system} as well as the unknown noise $n_i(\cdot)$ in the velocity feedback.

\section{Main Results} \label{sec:solution}

Let us first introduce the distance and orientation errors:
\begin{subequations} \label{eq:errors}
	
	\begin{align}
	e_k &\coloneqq \left\| p_{k_2}-p_{k_1} \right\|^2-d_{k,\text{des}}^2 \ \ \in\mathbb{R},  \label{eq:error e_k}\\
	\psi_k &\coloneqq \frac{1}{2} \text{tr}\Big[I_{3} - R^\top_{k,\text{des}} R^\top_{k_2} R_{k_1}  \Big] \ \ \in [0,2], \label{eq:error psi_k}
	\end{align}
\end{subequations}
$\forall k \in \mathcal{K}$. The fact that $\psi_k\in[0,2]$ is derived by using Proposition \ref{prop:R trace}. Regarding $e_k$, our goal is to guarantee $\lim_{t\to\infty} e_k(t) \to 0$ from all initial conditions satisfying \eqref{eq:at t=0}, while avoiding inter-agent collisions and connectivity losses among the initially connected agents specified by $\mathcal{E}$. Regarding $\psi_k$, we aim to guarantee the following: $1)$ $\lim_{t\to\infty}\psi_k(t) \to 0$, which according to Proposition \ref{prop:R trace} implies that $\lim_{t\to\infty} R_{k_2}(t)^\top R_{k_1}(t) = R_{k,\text{des}}$; $2)$ $\psi_k(t) < 2$, $\forall t\in\mathbb{R}_{\geq 0}$, since the configuration $\psi_k = 2$ is an undesired equilibrium, as will be clarified later.\footnote{It is well known that topological obstructions do not allow global stabilization on $\mathbb{SO}(3)$ with a continuous feedback control law (see \cite{lee_2017, lee_attitude_control_letters, lee2010control})} By invoking the properties of skew symmetric matrices of Section \ref{sec:preliminaries}, the errors \eqref{eq:errors} evolve according to the dynamics:
\begin{subequations} \label{eq:errors derivative}
	\begin{align}
	\dot{e}_k 
	& =2 (R^\top_{k_1}\widetilde{p}_{k_2,k_1})^\top (R^\top_{k_1}R_{k_2}v_{k_2,L}-v_{k_1,L}), \label{eq:error e_k dot} \\
	 \dot{\psi}_k &= \frac{1}{2}e_{R_k}^\top (R^\top_{k_1} R_{k_2}\omega_{k_2} - \omega_{k_1}), \label{eq:error psi_k dot} 
	\end{align}
\end{subequations} 
where $\widetilde{p}_{k_2,k_1} \coloneqq p_{k_2} - p_{k_1}$ and $e_{R_k}\coloneqq S^{-1}(R^\top_{k_1}R_{k_2}R_{k,\text{des}}-R^\top_{k,\text{des}}R^\top_{k_2}R_{k_1})$, $\forall k\in \mathcal{K}$. By employing Proposition \ref{prop:e_R frobenious}, we obtain $\|e_{R_k}\|^2 = \|R^\top_{k_2}R_{k_1}-R_{k,\text{des}}  \|^2_\text{F}(1 - \tfrac{1}{8} \|R^\top_{k_2}R_{k_1}-R_{k,\text{des}}\|^2_\text{F})$ as well as $\|R^\top_{k_2}R_{k_1}-R_{k,\text{des}}\|^2_\text{F}$ $=\text{tr}\Big[(R^\top_{k_2}R_{k_1}-R_{k,\text{des}})^\top(R^\top_{k_2}R_{k_1}-R_{k,\text{des}}) \Big]$ $= \text{tr}\left[2I_3 -2R^\top_{k,\text{des}}R^\top_{k_2}R_{k_1} \right]$ $= 4\psi_k \notag$.
Hence, it holds that:
\begin{equation} \label{eq:e_R_k and psi_k}
\|e_{R_k} \|^2 = 2\psi_k (2 - \psi_k),
\end{equation}
which implies that: $\|e_{R_k}\| = 0 \Rightarrow \psi_k = 0 \ \textit{or} \ \psi_k =2$, $\forall k\in\mathcal{M}$. The two configurations $\psi_k =0$ and $\psi_k = 2$ correspond to the desired and undesired equilibrium, respectively.

The concepts and techniques of prescribed performance control (see Section \ref{subsec:ppc}) are adapted in this work in order to: a) achieve predefined transient and steady state response for the distance and orientation errors $e_k$, $\psi_k$, $\forall k \in \mathcal{K}$, as well as ii) avoid the violation of the distance and connectivity constraints between initially neighboring agents, as presented in Section \ref{sec:prob_formulation}. The mathematical expressions of prescribed performance are given by the inequality objectives: 
\begin{subequations} \label{eq:ppc ineq}
	\begin{align}
	-C_{k,\text{col}} \rho_{e_k}(t) & < e_k(t) < C_{k,\text{con}} \rho_{e_k}(t), \\
	0 & \leq  \psi_{k}(t) < \rho_{\psi_k}(t) < 2, \label{eq:ppc ineq psi}
	\end{align}
\end{subequations}
$\forall k \in \mathcal{K}$, where $\rho_{e_k}:\mathbb{R}_{\geq 0} \to  \left[\tfrac{\rho_{\scr e_k,\infty}} {\max\{C_{k,\text{con}}, C_{k,\text{col}} \}},1\right]$,  $\rho_{\psi_k}:\mathbb{R}_{\geq 0}\to[\rho_{\scr \psi_k,\infty}, \rho_{\scr \psi_k,0}]$, with $\rho_{\psi_k}(t) \coloneqq (\rho_{\scr \psi_k,0} - \rho_{\scr \psi_k,\infty})e^{-l_{\psi_k} t} + \rho_{\scr \psi_k,\infty}$, $\rho_{e_k}(t)$ $\coloneqq \left[1 - \frac{\rho_{\scr e_k,\infty}} {\max\{C_{k,\text{con}}, C_{k,\text{col}} \}}\right]$ $e^{-l_{e_k} t}  + \frac{\rho_{\scr e_k,\infty}} {\max\{C_{k,\text{con}}, C_{k,\text{col}} \} }$, are designer-specified, smooth, bounded, and decreasing functions of time; the constants $l_{e_k}$, $l_{\psi_k}\in\mathbb{R}_{> 0}$, and $\rho_{\scr e_k,\infty}\in(0,\max\{C_{k,\text{con}}, C_{k,\text{col}} \})$, $\rho_{\scr \psi_k,\infty}\in (0,\rho_{\scr \psi_k,0})$, $\forall k \in \mathcal{K}$, incorporate the desired transient and steady state performance specifications respectively, as presented in Section \ref{subsec:ppc}, and $C_{k,\text{col}}$, $C_{k,\text{con}}\in\mathbb{R}_{>0},\forall k \in \mathcal{K}$, are associated with the distance and connectivity constraints. In particular, we select
\begin{equation} \label{eq:C_k}
C_{k,\text{col}} \coloneqq d^2_{k, \text{des}} - d^2_{k,\text{col}}, C_{k,\text{con}} \coloneqq d^2_{k,\text{con}} - d^2_{k, \text{des}},
\end{equation}
$\forall k \in \mathcal{K}$, which, since the desired formation is compatible with the constraints (i.e., $d_{k,\text{col}}<d_{k,\text{des}}<d_{k,\text{con}}, \forall k \in \mathcal{K}$), ensures that $C_{k,\text{col}},C_{k,\text{con}}\in\mathbb{R}_{>0},\forall k \in \mathcal{K}$, and consequently, in view of \eqref{eq:at t=0}, that: $-C_{k,\text{col}} \rho_{e_k}(0) < e_k(0) <\rho_{e_k}(0) C_{k,\text{con}}$, $\forall k \in \mathcal{K}$. Moreover, assuming that $\psi_k(0) < 2$, $\forall k\in \mathcal{K}$, by choosing: 
	\begin{equation}
	\rho_{\scr \psi_k,0} \coloneqq \rho_{\psi_k}(0) \in \Big(\psi_k(0),2\Big), \label{eq:rho_psi_0}
	\end{equation}
	it is also guaranteed that: $0 \leq \psi_k(0) < \rho_{\psi_k}(0)$,
$\forall k \in \mathcal{K}$. Hence, if we guarantee prescribed performance via \eqref{eq:ppc ineq}, by setting the steady state constants $\rho_{\scr e_k,\infty}, \rho_{\scr \psi_k,\infty}$ arbitrarily close to zero and by employing the decreasing property of $\rho_{e_k}(t),\rho_{\psi_k}(t),\forall k \in \mathcal{K}$, we guarantee practical convergence of the errors $e_k(t), \psi_k(t)$ to zero and we further obtain:
\begin{align} \label{eq:ppc goal}
-C_{k,\text{col}}  < e_k(t) < C_{k,\text{con}}, 0  \leq \psi_{k}(t) < \rho_{\psi_k}(t),
\end{align}
$\forall t\in\mathbb{R}_{\geq 0}$, which, owing to \eqref{eq:C_k}, implies: $d_{k,\text{col}} < \lVert p_{k_2}(t)-p_{k_1}(t)\rVert < d_{k,\text{con}}$, $\forall k \in \mathcal{K}, t \in \mathbb{R}_{\geq 0}$, providing, therefore, a solution to problem \ref{problem}. Moreover, note that the choice of $\rho_{\scr \psi_k,0}$ along with \eqref{eq:ppc goal} guarantee that $\psi_k(t) < 2$, $\forall t\in\mathbb{R}_{\geq 0}$ and the avoidance of the unstable singularity equilibrium.

In the sequel, we propose a decentralized control protocol that does not incorporate any information on the agents' dynamic model and guarantees \eqref{eq:ppc ineq} for all $t\in\mathbb{R}_{\geq 0}$. 
Given the errors $e_k, \psi_k$, we perform the following steps: 

\textbf{Step I-a}: Select the corresponding functions $\rho_{e_k}(t)$, $\rho_{\psi_k}(t)$ and positive parameters $C_{k,\text{con}}$, $C_{k,\text{col}}$, $k \in \mathcal{K}$, following \eqref{eq:ppc ineq}, \eqref{eq:rho_psi_0}, and \eqref{eq:C_k}, respectively, in order to incorporate the desired transient and steady state performance specifications as well as the distance and connectivity constraints, and define the normalized errors, $\forall k \in \mathcal{K}$,
	\begin{align}
	\xi_{e_k} \coloneqq \rho_{e_k}(t)^{-1}e_k, \  \xi_{\psi_k} \coloneqq  \rho_{\psi_k}(t)^{-1}\psi_k. \label{eq:ksi_k}
	\end{align}
\textbf{Step I-b}: Define the transformations $T_{e_k}:(-C_{k,\text{col}},$ $C_{k,\text{con}})$ $\to\mathbb{R}$, $k \in \mathcal{K}$,  and $T_{\psi}:[0,1)\to[0,\infty)$ by $T_{e_k}(x) \coloneqq \ln\Bigg(\frac{1+\tfrac{x}{C_{k,\text{col}}}}{1-\tfrac{x}{C_{k,\text{con}}}}\Bigg)$, $T_{\psi}(x) \coloneqq \ln\Big(\frac{1}{1 - x}\Big)$, $\forall k \in \mathcal{K}$, and the transformed error states, $\forall k \in \mathcal{K}$, 
\begin{equation} \label{eq:epsilon errors}
	\varepsilon_{e_k} \coloneqq T_{e_k}(\xi_{e_k}), \varepsilon_{\psi_k} \coloneqq T_{\psi}(\xi_{\psi_k}). 
\end{equation}
Next, we design the decentralized reference velocity vector for each agent $v_{i,\text{des}} \coloneqq [v_{i,L\text{des}}^\top, \omega^\top_{i,\text{des}}]^\top$ as
\begin{align}
&v_{i,\text{des}} = 
\begin{bmatrix}
v_{i,L\text{des}}\\ 
\omega_{i,\text{des}} 
\end{bmatrix} = \notag \\
&\hspace{-5mm} -\delta_i\begin{bmatrix}
2\sum\limits_{k\in\mathcal{M}} \alpha(i,k,R_{k_1},R_{k_2})\frac{r_{e_k}(\xi_{e_k})}{\rho_{e_k}(t)}\varepsilon_{e_k} R^\top_{k_1}\widetilde{p}_{k_2,k_1}  \\
\sum\limits_{k\in\mathcal{K}}  \alpha(i,k,R_{k_1},R_{k_2})\frac{r_{\psi}(\xi_{\psi_k})}{\rho_{\psi_k}(t)}e_{R_k}
\end{bmatrix},\hspace{-3mm}\label{eq:vel_i_des}
\end{align}
where $\delta_i\in\mathbb{R}_{>0}$ are positive gains, $\forall i\in\mathcal{N},$ $r_{e_k}:(-C_{k,\text{col}},C_{k,\text{con}})\to[1,\infty), r_{\psi}:[0,1)\to[1,\infty)$, with $r_{e_k}(x) \coloneqq \frac{\partial T_{e_k}(x)}{\partial x}$, $r_{\psi}(x)\coloneqq \frac{\partial T_{\psi}(x)}{\partial x}$,
 and $\alpha$ is defined as $\alpha(i,k,R_{k_1},R_{k_2}) = -I_3$, if $i$ is the tail of the $k$th edge ($i=k_1$), $\alpha(i,k,R_{k_1},R_{k_2}) = R^\top_{k_2} R_{k_1}$ if $i$ is the head of the $k$th edge ($i=k_2$), and $0$ otherwise. The assignment of the  head and tail in each edge can be done off-line according to the specified orientation of the graph, as mentioned in Section \ref{subsection: graph theory}.

\textbf{Step II-a}: Define for each agent the velocity errors $e_{v_i} \coloneqq [e_{v_i,1}^\top,\dots,e_{v_i,6}^\top]^\top \coloneqq \widetilde{v}_i - v_{i,\text{des}}$, $\forall i\in\mathcal{N}$, and design the decreasing performance functions as $\rho_{v_{i,\ell}}:\mathbb{R}_{\geq 0} \to  [\rho_{\scriptscriptstyle v^{0}_{i,\ell}}, \rho_{\scr v^{\infty}_{i,\ell}}]$, with $\rho_{v_{i,\ell}}(t)\coloneqq (\rho_{\scriptscriptstyle v^{0}_{i,\ell}} - \rho_{\scriptscriptstyle v^{\infty}_{i,\ell}})\exp(-l_{v_{i,\ell}}t) + \rho_{\scriptscriptstyle v^{\infty}_{i,\ell}}$, where the constants $\rho_{\scriptscriptstyle v^{0}_{i,\ell}}, \rho_{\scriptscriptstyle v^{\infty}_{i,\ell}}, l_{v_{i,\ell}}$ incorporate the desired transient and steady state specifications, with the design constraints $\rho_{\scriptscriptstyle v^{0}_{i,\ell}} > | e_{v_{i,\ell}}(0)|$, $ \rho_{\scriptscriptstyle v^{\infty}_{i,\ell}}\in(0,\rho_{\scriptscriptstyle v^{0}_{i,\ell}})$, $\forall \ell \in\{1,\dots,6\}$, $i\in\mathcal{N}$. The term $e_{v_{i,\ell}}(0)$ can be measured by each agent at $t=0$ directly after the calculation of $v_{i,\text{des}}(0)$. Moreover, define the normalized velocity errors

\begin{align}
\xi_{v_i} 
& \coloneqq  [\xi_{v_{i,1}},\dots,\xi_{v_{i,6}}]^\top  \coloneqq \rho_{v_i}(t)^{-1} e_{v_i}, \label{eq:ksi_i_v}
\end{align}
where $\rho_{v_i}(\cdot)\coloneqq \text{diag}\{ [\rho_{v_{i,\ell}}(\cdot)]_{\ell \in \{1,\dots,6\}}\}$, $\forall i\in\mathcal{N}$.


\textbf{Step II-b}: Define the transformation $T_v:(-1,1)\to\mathbb{R}$ as: $T_v(x)\coloneqq\ln\Big(\frac{1+x}{1-x}\Big)$, 
and the transformed error states 
\begin{align}
\hspace{-4mm}\varepsilon_{v_i} \coloneqq [\varepsilon_{v_i,1}, \dots, \varepsilon_{v_i,6}]^\top 
= 
[T_v(\xi_{v_i,1}), \dots, T_v(\xi_{v_i,6}) ]^\top\hspace{-2mm}, \hspace{-4mm} \label{eq:epsilon v}
\end{align}
Finally, 
design the decentralized control protocol for each agent $i\in\mathcal{N}$ as 
\begin{equation}
\hspace{-3mm} u_i \coloneqq -\gamma_i \left[\rho_{v_i}(t)\right]^{-1}\bar{r}_v(\xi_{v_i})\varepsilon_{v_i}, \label{eq:u_i}
\end{equation}
where $\bar{r}_v(\xi_{v_i}) \coloneqq \text{diag}\{ [r_v(\xi_{v_{i,\ell}})]_{\ell\in\{1,\dots,6\}}\}$ with $r_{v}:(-1,1)\to[1,\infty)$, $r_v(x)\coloneqq \frac{\partial T_v(x)}{\partial x}$, and  $\gamma_i \in\mathbb{R}_{>0}$ are positive gains, $\forall i\in\mathcal{N}$. 

\begin{remark}$[$\textbf{Control protocol intuition}$]$ Note that the selection of $C_{k,\text{col}}, C_{k,\text{con}}$ according to \eqref{eq:C_k} and of $\rho_{\psi_k}(t),\rho_{v_{i,\ell}}(t)$ such that
	$\rho_{\scr \psi_k,0}=\rho_{\psi_k}(0) \in (\psi_k(0),2), \rho_{\scr v^0_{i,\ell}}=\rho_{v_{i,\ell}}(0) > \lvert e_{v_{i,\ell}}(0)\rvert$ along with \eqref{eq:at t=0}, guarantee that $\xi_{e_k}(0)\in(C_{k,\text{col}},C_{k,\text{con}})$, $\psi_{k}(0)\in[0,2)$, $\xi_{\scr v_{i,\ell}}(0)\in(-1,1)$, $\forall k \in \mathcal{K}$, $\ell\in\{1,\dots,6\}$, $i\in\mathcal{N}$. The prescribed performance control technique enforces these normalized 
	errors $\xi_{e_k}(t), \xi_{\psi_k}(t)$ and $\xi_{v_{i,\ell}}(t)$ to remain strictly within the sets $(
	-C_{k,\text{col}},C_{k,\text{con}}), [0,2)$, and $(-1,1)$, respectively, $\forall k \in \mathcal{K}, \ell \in \{1,\dots,6\}, i\in\mathcal{N},t\geq0$, guaranteeing thus a solution to Problem \ref{problem}. It can be verified that this can be achieved by maintaining the boundedness of the modulated errors $\varepsilon_{e_k}(t), \varepsilon_{\psi_k}(t)$ and $\varepsilon_{v_i}(t)$ in a compact set, $\forall t\geq0$.
\end{remark}


\begin{remark} $[$\textbf{Arbitrarily fast convergence to} $\psi_k = 0$$]$
	The configurations where $\|e_{R_k}\| = 0 \Leftrightarrow \psi_k = 0$ or $\psi_k = 2$ are equilibrium configurations that result in $\omega_{k_1,\text{des}} = \omega_{k_2,\text{des}} = 0$, $\forall k\in\mathcal{K}$. If $\psi_k(0) = 2$, which is a local minima, the orientation formation specification for edge $k$ cannot be met, since the system becomes uncontrollable. This is an inherent property of stabilization in $\mathbb{SO}(3)$, and cannot be resolved with a purely continuous controller \cite{bhat2000topological}. Moreover, initial configurations $\psi_k(0)$ starting arbitrarily close to $2$ might take infinitely long to be stabilized at $\psi_k = 0$ with common continuous methodologies \cite{mayhew2011}. Note however, that the proposed control law guarantees convergence to $\psi_k = 0$ arbitrarily fast, given that $\psi_k(0) < 2$. More specifically, given the initial configuration $\psi_k(0) < 2$, we can always choose $\rho_{\scr \psi_k,0}$  such that $\psi_k(0) < \rho_{\scr \psi_k,0} < 2$, regardless of how close $\psi_k(0)$ is  to $2$. Then, as proved in the next section, the proposed control algorithm guarantees \eqref{eq:ppc ineq psi} and the transient and steady state performance of the evolution of $\psi_k(t)$ is determined solely by $\rho_{\psi_k}(t)$ and more specifically, its convergence rate is determined solely by the term $l_{\psi_k}$. It can be observed from the desired angular velocities $\omega_{i,\text{des}}$, designed in \eqref{eq:vel_i_des}, that close to the configuration $\psi_k(0) = 2$, the term $e_{R_k}(0)$, which is close  to zero (since $\psi_k(0) = 2 \Rightarrow \|e_{R_k}(0)\| = 0$), is compensated by the term $r_\psi(\xi_{\psi_k}(0)) = \frac{1}{1 - \xi_{\psi_k}(0)}$, which attains large values (since $\xi_{\psi_k}(0) = \frac{\psi_k(0)}{\rho_{\scr \psi_k,0}}$ is close to 1). In previous related approaches, the term $e_{R_k}(0)$ renders the control input arbitrarily small in configurations arbitrarily close to $\psi_k(0) = 2$, resulting thus in arbitrarily large stabilization time. Finally, note that potentially large values (but always bounded, as proved in the next section) for $\omega_{i,\text{des}}$ and hence $u_i$ due to the term $r_\psi(\xi_{\psi_k}(0))$ can be compensated by tuning the control gains $\delta_i$ and $\gamma_i$.
\end{remark}

\begin{remark} $[$\textbf{Decentralized manner, relative feedback, and robustness}$]$
	Notice by \eqref{eq:vel_i_des} and \eqref{eq:u_i} that the proposed control protocols are distributed in the sense
	that each agent uses only local \textit{relative} information to calculate its
	own signal. In that respect, regarding every edge $k$, the parameters $\rho_{\scr e_k,\infty}, \rho_{\scr \psi_k,\infty}, l_{e_k}, l_{\psi_k}$, as well as the sensing radii $s_j,\forall j\in \mathcal{N}_i(p(0))$, which are needed for the calculation of the performance functions $\rho_{e_k}(t), \rho_{\psi_k}(t)$, can be transmitted off-line to the agents $k_1,k_2\in\mathcal{N}$. In the same vein, regarding $\rho_{v_{i,\ell}}(\cdot)$, i.e., the constants $\rho_{\scr v^{\infty}_{i,\ell}}, l_{v_{i,\ell}}$ can be transmitted off-line to each agent $i$, which can also compute $\rho_{\scr v^0_{i,\ell}}$, given the initial velocity errors $e_{v_i}(0)$. Notice also from \eqref{eq:vel_i_des} that each agent $i$ uses only relative feedback with respect to its neighbors. In particular, for the calculation of $v_{i,L\text{des}}$, the tail of edge $k$, i.e., agent $k_1$, uses feedback of $R^\top_{k_1}(p_{k_2}-p_{k_1})$, and the head of edge $k$, i.e., agent $k_2$, uses feedback of $R^\top_{k_2}R_{k_1}R^\top_{k_1}(p_{k_2}-p_{k_1}) = R^\top_{k_2}(p_{k_2}-p_{k_1})$. Both of these terms are the relative inter-agent position difference expressed in the respective agent's local frames. For the calculation of $\omega_{i,\text{des}}$, agents $k_1$ and $k_2$ require feedback of the relative orientation $R^\top_{k_2}R_{k_1}$, as well as the signal $S^{-1}(R^\top_{k_1}R_{k_2}R_{k,\text{des}} - R^\top_{k,\text{des}}R^\top_{k_2}R_{k_1})$, which is a function of $R^\top_{k_2}R_{k_1}$. The aforementioned signals encode information related to the relative pose of each agent with respect to its neighbors, without the need for knowledge of a common global inertial frame. 	
	It should also be noted that the proposed
	control protocol \eqref{eq:u_i} depends exclusively on
	the velocity of each agent (expressed in the agent's local frame) and not on the velocity of its neighbors.
	Moreover, the proposed control law does not incorporate any prior knowledge of
	the model nonlinearities/disturbances, enhancing thus its robustness. 
	Nevertheless, note that the proposed protocol does not guarantee collision avoidance among the agents that are not initially connected. Formation control with collision avoidance among all the agents has only been achieved by using appropriately designed potential fields (e.g., \cite{chris_alex_cdc,dimarogonas2007application,tanner2005formation,tanner2012multiagent}), and it features several disadvantages, like simplified and known dynamics, excessive gain tuning, and non-global results. It is part of our future directions to extend the proposed scheme to account for collision avoidance among all the agents of the network.
	Finally, the proposed methodology results in a low complexity. Notice that no hard
	calculations (neither analytic nor numerical) are required to output the
	proposed control signal.
\end{remark}

\begin{remark} $[$\textbf{Construction of performance functions and gain tuning}$]$ \label{rem:pp functions and gain tuning}
	Regarding the construction of the performance functions, we stress that the
	desired performance specifications concerning the transient and steady state
	response as well as the distance and connectivity constraints are introduced
	in the proposed control schemes via $\rho_{e_k}(t), \rho_{\psi_k}(t)$ and
	$C_{k,\text{col}}, C_{k,\text{con}} $, $k \in \mathcal{K}$.
	In addition, the velocity performance functions $\rho_{v_{i,\ell}}(t)
	$, impose prescribed performance on the velocity errors $e_{v_i}=v_i-v_{i,\text{des}}$, $i\in\mathcal{N}$. In this respect, notice that $v_{i,\text{des}}$ acts as a reference signal for the corresponding velocities $v_{i}$, $i\in\mathcal{N}$. However, it should be stressed that
	although such performance specifications are not required (only the
	neighborhood position and orientation errors need to satisfy predefined transient and steady
	state performance specifications), their selection affects both the evolution
	of the errors within the corresponding performance envelopes as well
	as the control input characteristics (magnitude and rate).  More specifically, relaxing the convergence rate and the steady state limit of the velocity performance functions leads to increased oscillatory behavior within the prescribed
	performance region, which is improved when considering 	tighter performance functions, enlarging, however, the control effort both in magnitude and rate.
	Nevertheless, the
	only hard constraint attached to their definition is related to their initial
	values. Specifically, $\rho_{\scr \psi_k,0}=\rho_{\psi_k}(0)\in(\psi_k(0),2), \rho_{\scr v^0_{i,\ell}}=\rho_{v_{i,\ell}}(0)> | e_{v_{i,\ell}}(0)|$, $\forall k \in \mathcal{K}$, $\ell \in\{1,\dots,6\}$, $i\in\mathcal{N}$. In the same vein, as will be verified by the proof of Theorem \ref{thm:main theorem}, the actual transient- and steady-state performance of the closed loop system is solely determined by the performance functions  $\rho_{e_k}(t)$, $\rho_{\psi_k}(t)$, $\rho_{v_{i,\ell}}(t)$, and the constants $C_{k,\text{col}}$, $C_{k,\text{con}}$,  $k\in\mathcal{K}, \ell\in\{1,\dots,6\}, i\in\mathcal{N}$, without requiring any tuning of the gains $\delta_i, \gamma_i$, $i\in\mathcal{N}$. It should be noted, however, that their selection affects the control input characteristics and the state trajectory in the prescribed performance area. In particular, decreasing the gain values leads to increased oscillatory behavior within the prescribed performance area, which is improved when
	adopting higher values, enlarging, however, the magnitude and rate of the control input. Fine gain tuning is also needed in cases where the control input's magnitude and rate need to be bounded by pre-specified saturation values, since, although the proposed methodology yields bounded control inputs, it does not guarantee explicit bounds. In such cases, gain tuning might be needed to guarantee that the magnitude and rate of the control input do not exceed these values. A detailed analysis regarding the acquirement of such bounds is found in \cite{alex_chris_ppc_formation_ifac}.
\end{remark}

\begin{remark} $[$\textbf{Formation rigidity}$]$ \label{rem:rigidity}
Note that the desired distance and orientation formation defined in this work is not ``rigid", in the sense that the agents can achieve it under more than one relative configurations. This contrasts with certain works in the related literature, where the desired formation can be visualized as a fixed geometric shape in the configuration space (see, e.g., \cite{anderson_yu_fidan_hendrickx_2008, krick_broucke_francis_2009, dorfler_francis_2010, oh_ahn_2011e}). 
\end{remark}

\subsection{Stability Analysis}

In this section we provide the main result of this paper, which is summarized in the following theorem.
\begin{thm} \label{thm:main theorem}
	Consider the multi-agent system described by the dynamics \eqref{eq:system_MAS}, under a static tree sensing graph $\mathcal{G}$, aiming at establishing a formation described by the desired offsets $d_{k,\text{des}}\in(d_{k,\text{col}},d_{k,\text{con}})$ and $R_{k,\text{des}}$, $\forall k\in\mathcal{K}$, while satisfying the distance and connectivity constraints between initially neighboring agents, represented by $d_{k,\text{col}}$ and $d_{k,\text{con}}$, $\forall k\in \mathcal{K}$. Then, the control protocol \eqref{eq:ksi_k}-\eqref{eq:u_i} guarantees the prescribed transient and steady-state performance  $-C_{k,\text{col}} \rho_{e_k}(t) < e_k(t) < C_{k,\text{con}} \rho_{e_k}(t)$,
	$0 \leq \psi_{k}(t) < \rho_{\psi_k}(t)$,
	$\forall k \in \mathcal{K}$, $t\in\mathbb{R}_{\geq 0}$, under all initial conditions satisfying $\psi_k(0) < 2$, $\forall k\in\mathcal{K}$ and \eqref{eq:at t=0}, providing thus a solution to Problem \ref{problem}. 
\end{thm}

\begin{pf}
We start by defining some vector and matrix forms of the introduced signals and functions: $e \coloneqq [e_1,\dots,e_K]^\top, \psi$ $\coloneqq [\psi_1,\dots,\psi_K]^\top$, $e_R$ $\coloneqq [ e_{R_1}^\top, \dots, e_{R_K}^\top]^\top$, $e_v \coloneqq [ e_{v_1}^\top, \dots$, $e_{v_N}^\top]^\top$, $\xi_a$ $\coloneqq [\xi_{a_1},\dots,\xi_{a_K}]^\top$, $\xi_v \coloneqq [\xi_{v_1}^\top,\dots, \xi_{v_N}^\top]^\top$, $\varepsilon_e$ $\coloneqq [\varepsilon_{e_1},\dots$, $\varepsilon_{e_K}]^\top$, $\varepsilon_\psi \coloneqq [\varepsilon_{\psi_1},\dots,\varepsilon_{\psi_K}]^\top,$ $\varepsilon_v \coloneqq [\varepsilon^\top_{v_1},\dots,\varepsilon^\top_{v_N}]^\top,\widetilde{p} \coloneqq [\widetilde{p}^\top_{1_2,1_1},\dots,\widetilde{p}^\top_{K_2,K_1}]^\top$,  $v_L\coloneqq [v_{1,L}^\top,\dots,v_{N,L}^\top]^\top,v_{L\text{des}}\coloneqq [v_{1,L\text{des}}^\top,\dots,v_{N,L\text{des}}^\top]^\top$, $\omega \coloneqq [\omega_1^\top,\dots,\omega_N^\top]^\top, \omega_{\text{des}}$ $\coloneqq [\omega_{1,\text{des}}^\top,\dots,\omega_{N,\text{des}}^\top]^\top$, $v_{\text{des}} \coloneqq [v_{1,\text{des}}^\top, \dots, v_{N,\text{des}}^\top]^\top, \rho_a(t)$ $\coloneqq \text{diag}\{[\rho_{a_k}(t)]_{k\in \mathcal{K}}\},$ $\rho_v(t) \coloneqq \text{diag}\{[\rho_{v_i}(t)]_{i\in\mathcal{N}}\}$, $r_e(\xi_e) \coloneqq \text{diag}\{ [r_{e_k}(\xi_{e_k})]_{k\in \mathcal{K}}\}$, $\Sigma_e(\xi_e,t)$ $\coloneqq r_e(\xi_e)\rho_e(t)^{-1}$, $\widetilde{r}_\psi(\xi_\psi) \coloneqq \text{diag}\{ [r_{\psi}(\xi_{\psi_k})]_{k\in \mathcal{K}}\}, \Sigma_\psi(\xi_\psi$, $t)$ $\coloneqq \widetilde{r}_\psi(\xi_\psi) \\ \rho_\psi(t)^{-1}$, $\widetilde{r}_v(\xi_v)$ $\coloneqq \text{diag}\{ [\bar{r}_{v}(\xi_{v_i})]_{i\in\mathcal{N}} \}$, $\Sigma_v(\xi_v,t)$ $\coloneqq \widetilde{r}_v(\xi_v)\rho_v(t)^{-1}$, where $a\in\{e,\psi\}$.
 
With the introduced notation, \eqref{eq:errors derivative} can be written in vector form as 

\begin{subequations}  \label{eq:e_deriv}
	\begin{align}
	&\hspace{-2mm}\dot{e} = 
	\mathbb{F}_p(\widetilde{p})^\top \hat{R} D_R(R,\mathcal{G})^\top v_L, \label{eq:e_p_deriv} \\ 
	\dot{\psi} &=
	\mathbb{F}_{R}(e_R)^\top D_R(R,\mathcal{G})^\top\omega,
	\label{eq:e_q_deriv} 
	\end{align}
\end{subequations}
where $\hat{R} \coloneqq \text{diag}\{[R_{k_1}]_{k\in\mathcal{K}}\}\in\mathbb{R}^{3K\times 3K}$, $\mathbb{F}_p(\widetilde{p}) \coloneqq
2 \begin{bmatrix}
\widetilde{p}_{1_2,1_1}  & \dots & 0_{3\times 1} \\
\vdots & \ddots & \vdots\\
0_{3\times 1} &  \dots  & \widetilde{p}_{K_2,K_1}
\end{bmatrix} \in \mathbb{R}^{3K\times K}$,
$\mathbb{F}_R(e_R)  \coloneqq \\
 \frac{1}{2}\begin{bmatrix}
e_{R_1} & \dots & 0_{3\times 1} \\
 \vdots & \ddots & \vdots\\	
 0_{3\times 1} &  \dots  & e_{R_K} \\
 \end{bmatrix} \in \mathbb{R}^{3K\times K}$, and $D_R\in \mathbb{R}^{3N}\times\mathbb{R}^{3K}$ is the \emph{orientation incidence matrix} of the graph: 
\begin{align} 
& D_R(R,\mathcal{G}) \coloneqq  \bar{R}^\top\left[ D(\mathcal{G})\otimes I_3 \right] \hat{R} , \label{eq:incidence_D_R}
\end{align}
with $\bar{R}\coloneqq\text{diag}\{[R_i]_{i\in\mathcal{N}}\}\in\mathbb{R}^{3N\times3N}$, and $D(\mathcal{G})$ is the incidence matrix of the graph. 
The terms $\bar{R}$ and $\hat{R} $ in $D_R(R,\mathcal{G})$ correspond to the block diagonal matrix with the agents' rotation matrices along the main block diagonal, and the block diagonal matrix with the rotation matrix of each edge's tail along the main block diagonal, respectively. These two terms have motivated the incorporation of the terms $\alpha(\cdot)$ in the desired velocities $v_{i,\text{des}}$ designed in \eqref{eq:vel_i_des}, since,  as shown next, the vector form $v_{\text{des}}$ yields the orientation incidence matrix $D_{R}(R,\mathcal{G})$.  

The desired velocities \eqref{eq:vel_i_des} and control inputs \eqref{eq:u_i} can be written in vector form as 
\begin{subequations} \label{eq:control design vectors}
	\begin{align}
	v_{L\text{des}} &= -\Delta D_R(R,\mathcal{G})\hat{R} ^\top\mathbb{F}_p(\widetilde{p})\Sigma_e(\xi_e,t)\varepsilon_e, \label{eq:control design p vectors} \\
	\omega_\text{des} &= -\Delta D_R(R,\mathcal{G}) \left[ \Sigma_\psi(\xi_\psi,t) \otimes I_3 \right] e_R, \label{eq:control design R vectors}\\
	u &= -\Gamma \ \Sigma_v(\xi_v,t)\varepsilon_v, \label{eq:control design u vectors}
	\end{align}
\end{subequations}
where $\Delta \coloneqq \text{diag}\{[\delta_iI_3]_{i\in\mathcal{N}}\}\in\mathbb{R}^{3N\times3N}$ and  $\Gamma \coloneqq \text{diag}\{[\gamma_iI_6]_{i\in\mathcal{N}} \}\in\mathbb{R}^{6N\times6N}$. Note from \eqref{eq:control design u vectors} and \eqref{eq:ksi_k}, \eqref{eq:ksi_i_v}, \eqref{eq:epsilon errors}, \eqref{eq:epsilon v} that $u$ can be expressed as a function of the states $u(x,v,t)$.  Hence, the closed loop system can be written as $\dot{x} =  h_x(x,v)$, $\dot{v} =h_v(x,v,t) \coloneqq -M^{-1}\{ C(v)v + g(x) + w(x,v,t) - u(x,v,t) \}$,
and by defining $z \coloneqq (x,v)\in\mathbb{SE}(3)^{N}\times\mathbb{R}^{6N}$:
\begin{equation}
	\dot{z} = h(z,t) \coloneqq (h_x(z), h_v(z,t)). \label{eq:closed loop vector}
\end{equation}
Next, define the set $\Omega \coloneqq \{  (x,v,t)\in\mathbb{SE}(3)^N\times\mathbb{R}^{6N}\times\mathbb{R}_{\geq 0} :
 \xi_{e_k}(p_{k_1},p_{k_2},t) \in (-C_{k,\text{col}}, C_{k,\text{con}}),  \xi_{\psi_k}(R_{k_1},R_{k_2},t)$ $< 1, 
\xi_{v_i}(x,v_i,t)\in(-1,1)^6, \forall k\in \mathcal{K}\}$,
where we have expressed $\xi_{e_k}$, $\xi_{\psi_k}$, $\xi_{v_i}$ from \eqref{eq:ksi_k}, \eqref{eq:ksi_i_v} as a function of the states. It can be verified that the set $\Omega$ is open due to the continuity of the operators $\xi_{e_k}(\cdot), \xi_{\psi_k}(\cdot), \xi_{v_i}(\cdot)$ and nonempty, due to \eqref{eq:C_k}. Our goal here is to prove firstly that \eqref{eq:closed loop vector} has a unique and maximal solution $(z(t),t)$ in $\Omega$ and then that this solution stays in a compact subset of $\Omega$.

It can be verified that the function $h:\Omega  \to  \mathbb{T}_{R_1}\times\cdots\times\mathbb{T}_{R_N} \times\mathbb{R}^{6N}$ is (a) continuous in $t$ for each fixed $(x,v)\in\{ (x,v)\in \mathbb{SE}(3)^N\times\mathbb{R}^{6N}: (x,v,t)\in\Omega\}$, and (b) continuous and locally lipschitz in $(x,v)$ for each fixed $t\in\mathbb{R}_{\geq 0}$. Therefore, the conditions of Theorem \ref{thm:ode solution} are satisfied and hence, we conclude the existence of a unique and maximal solution of \eqref{eq:closed loop vector} for a timed interval $[0,t_{\max})$, with $t_{\max} >0$ such that $(z(t),t)\in\Omega$, $\forall t\in[0,t_{\max})$. This implies that 
\begin{subequations} \label{eq:ksi tau_max}
\begin{align}
	\xi_{e_k}(t) & = \rho_{e_k}(t)^{-1}e_k(t)\in(-1,1) \label{eq:ksi e tau_max}, \\
	\xi_{\psi_k}(t) & = \rho_{\psi_k}(t)^{-1}\psi_k(t) < 1, \label{eq:ksi psi tau_max} \\
	\xi_{v_i}(t) & = \rho_{v_i}(t)^{-1}e_{v_i}(t)\in (-1,1)^6, \label{eq:ksi v tau_max} 
\end{align}
\end{subequations}
$\forall k\in \mathcal{K}$, $i\in\mathcal{N}$,  $t\in[0,t_{\max})$. Therefore, the signals $e_k(t), \psi_k(t), e_{v_i}(t)$ are bounded for all $t\in[0,t_{\max})$. In the following, we aim to show that the solution $(z(t),t)$ is bounded in a compact subset of $\Omega$ and hence, by employing Theorem \ref{thm:forward_completeness}, that $t_{\max} = \infty$. 
 
Consider the positive definite function $V_e \coloneqq \tfrac{1}{2}\|\varepsilon_e\|^2$, which is well defined for $t\in[0,t_{\max})$, due to \eqref{eq:ksi e tau_max}. By differentiating $V_e$, 
we obtain $\dot{V}_e = \varepsilon^\top_e \Sigma_e(\xi_e,t) \{-  \dot{\rho}_e(t)\xi_e$ $+ \mathbb{F}_p(\widetilde{p})^\top \hat{R} D_R(R,\mathcal{G})^\top v_{L}  \}$, which, by substituting $v_L = \widetilde{v}_L - n_{p}(x,t) = e_{v_p} + v_{L\text{des}} - n_p(x,t)$ and \eqref{eq:e_deriv},  becomes:

\small
\begin{align*}
&\dot{V}_e = -\varepsilon^\top_e \Sigma_e(\xi_e,t)\mathbb{F}_p(\widetilde{p})^\top \widetilde{D}(\mathcal{G})\mathbb{F}_p(\widetilde{p})\Sigma_e(\xi_e,t)\varepsilon_e+ \\
&\varepsilon^\top_e \Sigma_e(\xi_e,t) \Big[ \mathbb{F}_p(\widetilde{p})^\top \hat{R} D_R(R,\mathcal{G})^\top(e_{v_p} - n_p(x,t))  -\dot{\rho}_e(t)\xi_e \Big],
\end{align*}
\normalsize
where $\widetilde{D}(\mathcal{G})$ $\coloneqq$ $\hat{R} D_R(R,\mathcal{G})^\top D_R(R,\mathcal{G}) \hat{R} ^\top$ $=$ $(D(\mathcal{G})^\top\otimes I_3)$ $\Delta$ $(D(\mathcal{G})$ $\otimes I_3) \in\mathbb{R}^{3K\times 3K}$ (by employing \eqref{eq:incidence_D_R}), and $e_{v_p}$, $n_p(x,t)$ are the linear parts of $e_v$ and $n(x,t)$ (i.e., the stack vector of the first three components of every $e_{v_i}$, $n_i(x_i,t)$), respectively. Note first that, due to \eqref{eq:ksi v tau_max}, the function $e_{v_p}(t)$ is bounded for all $t\in[0,t_{\max})$. Moreover, note that \eqref{eq:ksi e tau_max} implies that $0<d_{k,\text{col}} < \|p_{k_1}(t)-p_{k_2}(t) \| < d_{k,\text{con}}$, $\forall t\in[0,t_{\max})$. Therefore, it holds that $\text{rank} (\mathbb{F}_p(\widetilde{p}(t))) = K$, $\forall t\in[0,t_{\max})$. In addition, since $G$ is a connected tree graph and $\delta_i\in\mathbb{R}_{>0}$, $\forall i\in\mathcal{N}$, $\widetilde{D}(\mathcal{G})$ is positive definite and $\text{rank}(\widetilde{D}(\mathcal{G})) = 3K$. Hence, we conclude that $\text{rank}\Big( [\mathbb{F}_p(\widetilde{p}(t))]^\top\widetilde{D}(\mathcal{G})\mathbb{F}_p(\widetilde{p}(t))\Big) = K$ and the positive definiteness of $[\mathbb{F}_p(\widetilde{p}(t))]^\top\widetilde{D}(\mathcal{G})\mathbb{F}_p(p(t))$, $\forall t\in[0,t_{\max})$ is deduced. In addition, since $\|p_{k_2}(t) - p_{k_1}(t)\| < d_{k,\text{con}}$, we also conclude that the term $\mathbb{F}_p(\widetilde{p})^\top\hat{R} D_R(R,\mathcal{G})^\top$ is upper bounded, $\forall t\in[0,t_{\max})$. Finally, $\dot{\rho}_e(t)$ and $n_p(x,t)$ are bounded by definition and assumption, respectively, $\forall x\in\mathbb{SE}(3)^N, t\in\mathbb{R}_{\geq 0}$. We obtain $\dot{V}_e \leq  - \underline{\lambda}_{\scr \widetilde{D}} \| \Sigma_e(\xi_e,t)\varepsilon_e \| \left[\| \Sigma_e(\xi_e,t)\varepsilon_e \| - \frac{\bar{B}_e}{\underline{\lambda}_{\scr \widetilde{D}}} \right]$,
$\forall t\in[0,t_{\max})$, 
where 
\begin{align*}
	\underline{\lambda}_{\scr \widetilde{D}} & \coloneqq \min\limits_{\scr p(t),t\in[t_0,t_{\max})}\Big\{\lambda_{\min}\Big(\mathbb{F}_p(\widetilde{p}(t))^\top\widetilde{D}(\mathcal{G})\mathbb{F}_p(\widetilde{p}(t))\Big)\Big\} \\
	& \geq d^2_{k,\text{col}}\lambda_{\min}(\widetilde{D}(\mathcal{G})) > 0,
\end{align*}
and $\bar{B}_e$ is a positive constant, independent of $t_{\max}$, satisfying
$\bar{B}_e \geq \|\mathbb{F}_p(\widetilde{p})^\top\hat{R} D_R(R,\mathcal{G})^\top$ $(e_{v_p}(t) - n_p(x,t)) - \dot{\rho}_e(t)\xi_e(t) \|, \forall t\in[0,t_{\max})$. Note that, in view of the aforementioned discussion, $\bar{B}_e$ is finite. 

Hence, we conclude that $\dot{V}_e < 0 \Leftrightarrow \| \Sigma_e(\xi_e,t)\varepsilon_e \| > \frac{\bar{B}_e}{\underline{\lambda}_{\scr \widetilde{D}}}$. It holds that $r_{e_k}(x)$ $=$ $\frac{\partial T_{e_k}(x)}{\partial x}$ $=$ $\frac{\frac{1}{C_{k,\text{col}}}+\frac{1}{C_{k,\text{con}}}}{\left(1 + \frac{x}{C_{k,\text{col}}}\right)\left(1-\frac{x}{C_{k,\text{con}}}\right)}$  
$>$ $\frac{1}{C_{k,\text{col}}} + \frac{1}{C_{k,\text{con}}}$,
$\forall x\in(-C_{k,\text{col}},C_{k,\text{con}})$, and $\rho_{e_k}(t) \leq 1, \forall t\in\mathbb{R}_{\geq 0}$, $k\in \mathcal{K}$, and thus we conclude that $\| \Sigma_e(\xi_e(t),t)\varepsilon_e(t) \|= \sqrt{\sum_{k\in\mathcal{K}} \frac{[r_{e_k}(\xi_{e_k}(t))]^2}{[\rho_{e_k}(t)]^2}[\varepsilon_{e_k}(t)]^2} \geq \bar{C}\|\varepsilon_e(t)\|$, $\forall t\in[0,t_{\max})$, where $\bar{C} \coloneqq \max\limits_{k\in\mathcal{K}}\left\{ \frac{C_{k,\text{col}}+C_{k,\text{con}}}{C_{k,\text{col}}C_{k,\text{con}}}\right\}$.  Hence, we conclude that 
$\dot{V}_e(\varepsilon_e) < 0$, $\forall \|\varepsilon_e\| \geq 
\tfrac{\bar{B}_e}{\underline{\lambda}_{\scr \widetilde{D}}\bar{C}}$, $\forall t\in[0,t_{\max})$. Therefore, by invoking Theorem $4.8$ in \cite{khalil_nonlinear_systems} we conclude that 
\begin{equation} \label{eq:bar epsilon e}
	\|\varepsilon_e(t) \| \leq \bar{\varepsilon}_e \coloneqq \max\left\{ \varepsilon_e(0), \tfrac{\bar{B}_e}{\underline{\lambda}_{\scr \widetilde{D}}\bar{C}} \right\},
\end{equation}
$t\in[0,t_{\max})$, and by taking the inverse logarithm function:
\begin{equation} \label{eq:bar ksi e}
	- C_{k,\text{col}} < -\underline{\xi}_{e} \leq \xi_{e_k}(t) \leq \bar{\xi}_{e} <  C_{k,\text{con}},
\end{equation}
$\forall t\in[0,t_{\max})$, where $\bar{\xi}_e \coloneqq \tfrac{\exp(\bar{\varepsilon}_e)-1}{\exp(\bar{\varepsilon}_e)+1}C_{k,\text{con}}$, and $\underline{\xi}_e \coloneqq \tfrac{\exp(-\bar{\varepsilon}_e)-1}{\exp(-\bar{\varepsilon}_e)+1}C_{k,\text{con}}$. Note that $\varepsilon_e(0)$ is finite due to the assumption $d_{k,\text{col}} < \|p_{k_2}(0)-p_{k_1}(0) \| < d_{k,\text{con}}$. Therefore, since $\underline{\lambda}_{\widetilde{D}}$ is strictly positive and $\bar{B}_e$ is also finite, $\bar{\varepsilon}_e$ is well defined. Hence, \eqref{eq:bar epsilon e} and \eqref{eq:bar ksi e} imply the boundedness of $\varepsilon_{e_k}(t)$, $r_{e_k}(\xi_{e_k}(t)$, $\widetilde{p}(t)$, and $p(t)$ in compact sets, $\forall k\in \mathcal{K}$, and therefore, through \eqref{eq:vel_i_des}, the boundedness of $v_{i,L\text{des}}(t)$, $\forall i\in\mathcal{N}$, $t\in[0,t_{\max})$. 

Similarly, consider the positive definite function $V_\psi = 2\sum_{k\in \mathcal{K}}\varepsilon_{\psi_k}$, whose derivative is $\dot{V}_\psi=$ $2\sum_{k\in \mathcal{K}} \frac{r_{\psi}(\xi_{\psi_k})}{\rho_{\psi_k}(t)}(\dot{\psi}_{k}$ $- \dot{\rho}_{\psi_k}\xi_{\psi_k})$. After substituting \eqref{eq:error psi_k dot}, \eqref{eq:e_deriv},  we obtain 
\begin{align*}
&\dot{V}_\psi = -2\sum\limits_{k\in \mathcal{K}}\frac{r_{\psi}(\xi_{\psi_k})}{\rho_{\psi_k}(t)}\dot{\rho}_{\psi_k}(t)\xi_{\psi_k} + \\
& e_R^\top \left[ \Sigma_\psi(\xi_\psi,t)\otimes I_3 \right] D_R(R,\mathcal{G})^\top \big[  \omega_\text{des} + e_{v_R}- n_R(x,t) \big],
\end{align*}
where  $e_{v_R}$ and $n_R(x,t)$ are the angular parts of $e_v$ and $n(x,t)$ (i.e., the stack vector of the last three components of every $e_{v_i}$, $n_i(x,t)$), respectively. By substituting \eqref{eq:control design R vectors} and defining $\widetilde{\Sigma}_\psi(\xi_\psi,t) \coloneqq \Sigma_\psi(\xi_\psi,t)\otimes I_3\in\mathbb{R}^{3K \times 3K}$, $\widetilde{D}_R(R,\mathcal{G}) \coloneqq D_R(R,\mathcal{G})^\top \Delta D_R(R,\mathcal{G})\in\mathbb{R}^{3K \times 3K}$, we obtain 
\begin{align*}
	\dot{V}_\psi = \ & -e_R^\top \widetilde{\Sigma}_\psi(\xi_\psi,t)\widetilde{D}_R(R,\mathcal{G})\widetilde{\Sigma}_\psi(\xi_\psi,t)e_R   \notag\\
	\ &+e_R^\top\widetilde{\Sigma}_\psi(\xi_\psi,t)D_R(R,\mathcal{G})^\top \left[ e_{v_R} - n_R(x,t) \right] \notag\\
	\ & - 2\sum\limits_{k\in\mathcal{K}}\frac{r_{\psi}(\xi_{\psi_k})}{\rho_{\psi_k}(t)}\dot{\rho}_{\psi_k}(t)\xi_{\psi_k}. 
\end{align*}
According to \eqref{eq:incidence_D_R}, $D_R(R,\mathcal{G})$ $=$ 
$\bar{R}^\top$ $\left[ D(\mathcal{G})\otimes I_3 \right]$ $\hat{R} $. Since $\bar{R}$ and $\hat{R} $ are rotation (and thus unitary) matrices, the singular values of $D_R(R,\mathcal{G})$ are identical to the ones of $D(\mathcal{G})$, and hence $\lambda_{\min}(\widetilde{D}_R(R,\mathcal{G})) = \lambda_{\min}(\widetilde{D}(\mathcal{G})) > 0$. Indeed, let $D(\mathcal{G})\otimes I_3 = U\Sigma_D V^\top$ be a singular value decomposition of $D(\mathcal{G})\otimes I_3$, where $U$, $V$ are unitary matrices, and $\Sigma_D$ is a diagonal matrix containing the singular values of $D(\mathcal{G})\otimes I_3$. Then 
$ D_R(R, \mathcal{G}) = \bar{R}^\top U \Sigma_D V^\top \hat{R} 
=  \widetilde{U}\Sigma_D \widetilde{V}^\top$
where $\widetilde{U} \coloneqq \bar{R}^\top U$, and $\widetilde{V} = \hat{R} ^\top V$ are unitary matrices (being products of unitary matrices). Thus, $\widetilde{U}\Sigma_D\widetilde{V}^\top$ is the singular value decomposition of $D_R(R,\mathcal{G})$, and hence its singular values are the diagonal values of $\Sigma_D$. By further defining $\beta \coloneqq$ $[\beta_1^\top,\dots,\beta_K^\top]^\top$ $\coloneqq$ $D_R(R,\mathcal{G})^\top (e_{v_R} - n_R(x,t))\in\mathbb{R}^{3M}$, with $\beta_k\in\mathbb{R}^3$, $\forall k\in\mathcal{K}$, $\dot{V}_\psi$ becomes 
\begin{align*}
	\dot{V}_\psi\leq& -\lambda_{\min}( \widetilde{D}(\mathcal{G}) )\| \widetilde{\Sigma}_\psi(\xi_\psi,t)e_R \|^2 \notag \\
	&\hspace{-10mm} + \sum\limits_{k\in\mathcal{K}}\frac{r_{\psi}(\xi_{\psi_k})}{\rho_{\psi_k}(t)} (e_{R_k})^\top \beta_k  - 2\sum\limits_{k\in\mathcal{K}}\frac{r_{\psi}(\xi_{\psi_k})}{\rho_{\psi_k}(t)}\dot{\rho}_{\psi_k}(t)\xi_{\psi_k}.
\end{align*}
Note that, by construction, $\xi_{\psi_k} \geq 0$, $\forall k\in\mathcal{K}$, and $r_\psi(x) = \frac{\partial T_\psi(x)}{\partial x} = \frac{1}{1-x} > 1, \forall x < 1$. Hence, in view of \eqref{eq:ksi psi tau_max}, we conclude that $r_\psi(\xi_{\psi_k}(t)) > 1$, $\forall t\in[0,t_{\max})$. By noting also that $\dot{\rho}_{\psi_k}(t) < 0,\forall t\in\mathbb{R}_{\geq 0}$ and after substituting \eqref{eq:e_R_k and psi_k}, $\dot{V}_\psi$ becomes
\begin{align*}
	\dot{V}_\psi(\varepsilon_\psi) \leq & -\lambda_{\min}( \widetilde{D}(\mathcal{G}) )\sum\limits_{k\in \mathcal{K}}\Bigg[\frac{r_{\psi}(\xi_{\psi_k})}{\rho_{\psi_k}(t)}\Bigg]^2\| e_{R_k} \|^2 \\ 
	& + \bar{B}_{\psi_1}\sum\limits_{k\in \mathcal{K}}\frac{r_{\psi}(\xi_{\psi_k})}{\rho_{\psi_k}(t)} \|e_{R_k}\| \\
	& + 2\max\limits_{k\in \mathcal{K}}\{ l_{\psi_k}(\rho_{\scr \psi_k,0} - \rho_{\scr \psi_k,\infty}) \}\sum\limits_{k\in \mathcal{K}}\frac{r_{\psi}(\xi_{\psi_k})}{\rho_{\psi_k}(t)}\xi_{\psi_k},
\end{align*}
where $\bar{B}_{\psi_1}$ is a positive constant, independent of $t_{\max}$, satisfying $\bar{B}_{\psi_1} \geq \max_{k\in \mathcal{K}}\{\| \beta_k(t) \|\}$, $\forall t\in[0,t_{\max})$. Note that $\bar{B}_{\psi_1}$ is finite, $\forall t\in[0,t_{\max})$, due to \eqref{eq:ksi psi tau_max} and the boundedness of the noise signals $n(x,t)$. 

From \eqref{eq:ksi psi tau_max} and the definition of $\psi_k$, we conclude that $0\leq \psi_k(t) < \rho_{\psi_k}(t) \leq \rho_{\scr \psi_k,0} < 2$, and hence $2-\psi_k(t) \geq 2 - \rho_{\scr \psi_k,0}=: \underline{\rho}_k > 0$ $\forall t\in[0,t_{\max})$, $k \in \mathcal{K}$. Moreover, by noticing that $2-\psi_k\leq 2$, $\rho_{\psi_k}(t) \leq \rho_{\scr \psi_k,0}$, and $\psi_k = \xi_{\psi_k}\rho_{\psi_k}(t)$, $\forall k\in \mathcal{K}$, $\dot{V}_\psi$ becomes 
\begin{align*}
	&\dot{V}_\psi \leq -\widetilde{\mu} \sum\limits_{k\in \mathcal{K}}\left[ r_{\psi}(\xi_{\psi_k}) \right]^2 \xi_{\psi_k} \\
	& \hspace{8mm} + \frac{2\bar{B}_{\psi_1}}{\max\limits_{k\in \mathcal{K}}\{\sqrt{\rho_{\scr \psi_k,0}}\}}\sum\limits_{k\in \mathcal{K}} r_{\psi}(\xi_{\psi_k}) \sqrt{\xi_{\psi_k}}
\end{align*}
\begin{align*}
	& \hspace{8mm} + 2\max\limits_{k\in \mathcal{K}}\left\{ \frac{l_{\psi_k}(\rho_{\scr \psi_k,0} - \rho_{\scr \psi_k,\infty})}{\rho_{\scr \psi_k,0}} \right\} \sum\limits_{k\in \mathcal{K}}r_{\psi}(\xi_{\psi_k})\xi_{\psi_k},
\end{align*} 
where $\widetilde{\mu} \coloneqq \frac{2\lambda_{\min}(\widetilde{D}(\mathcal{G}))\min_{k\in \mathcal{K}}\{\underline{\rho}_k\}}{\max_{k\in \mathcal{K}}\{\rho_{\scr \psi_k,0}\}}$.
In view of \eqref{eq:ksi psi tau_max}, it holds that $\xi_{\psi_k}(t) < \sqrt{\xi_{\psi_k}(t)}, \forall k\in \mathcal{K}$. By also employing $\sum_{k\in \mathcal{K}} r_{\psi_k}(\xi_{\psi_k})\sqrt{\xi_{\psi_k}} \leq \sqrt{K}\sqrt{\sum_{k\in \mathcal{K}} (r_{\psi_k}(\xi_{\psi_k}))^2\xi_{\psi_k}}$, we obtain 

\small
\begin{align*}
&\dot{V}_\psi \leq -\sqrt{\sum\limits_{k\in \mathcal{K}} \left[ r_{\psi}(\xi_{\psi_k}) \right]^2 \xi_{\psi_k}}  \Bigg\{\widetilde{\mu}\sqrt{\sum\limits_{k\in \mathcal{K}}\left[ r_{\psi_k}(\xi_{\psi_k}) \right]^2 \xi_{\psi_k}} - \bar{B}_\psi\Bigg\},
\end{align*} 
\normalsize
where 

\small
\begin{align*}
\bar{B}_\psi \coloneqq 2\sqrt{K}\Bigg[\frac{\bar{B}_{\psi_1}}{\max\limits_{k\in \mathcal{K}}\{\sqrt{\rho_{\scr \psi_k,0}}\}} + \max\limits_{k\in \mathcal{K}}\left\{ \frac{l_{\psi_k}(\rho_{\scr \psi_k,0} - \rho_{\scr \psi_k,\infty})}{\rho_{\scr \psi_k,0}} \right\}\Bigg].
\end{align*} 
\normalsize
Therefore, $\dot{V}_\psi < 0$ $\Leftrightarrow$ $\sqrt{\sum_{k\in \mathcal{K}} \left[r_{\psi}(\xi_{\psi_k}) \right]^2 \xi_{\psi_k}} > \tfrac{\bar{B}_\psi}{\widetilde{\mu}}$. From \eqref{eq:epsilon errors}, given $y = T_\psi(x)$, we obtain $\left[ r_{\psi}(x) \right]^2 x = \left[\frac{\partial T(x)}{\partial x}\right]^2 T^{-1}(y) = \frac{1}{(1-x)^2}T^{-1}(y)  =\frac{1}{\left[1-T^{-1}(y)\right]^2}T^{-1}(y) =   \exp(y)\left[\exp(y)-1\right]$, 
$\forall x\in[0,1)$. Therefore,
$ \left[r_{\psi}(\xi_{\psi_k})\right]^2\xi_{\psi_k}$ $= \exp(\varepsilon_{\psi_k})$ $\left[\exp(\varepsilon_{\psi_k})-1 \right]$, and
according to Prop. \ref{prop f(x)},  
\begin{align*}
	\sqrt{\sum_{k\in \mathcal{K}}\left[r_{\psi}(\xi_{\psi_k})\right]^2 \xi_{\psi_k}} & = \sqrt{\sum_{k\in \mathcal{K}} \exp(\varepsilon_{\psi_k})\left[ \exp(\varepsilon_{\psi_k})-1 \right]} \\
	& \geq \sqrt{\sum_{k\in \mathcal{K}} \varepsilon^2_{\psi_k}} = \| \varepsilon_{\psi}\|.
\end{align*}
Hence, we conclude that $\dot{V}_\psi < 0, \forall \| \varepsilon_{\psi}\| > \tfrac{\bar{B}_{\psi}}{\widetilde{\mu}}$. Therefore, 
\begin{align} \label{eq:bar epsilon psi}
	\| \varepsilon_\psi(t) \| \leq \bar{\varepsilon}_\psi \coloneqq \max\left\{\varepsilon_\psi(0), \tfrac{\bar{B}_\psi}{\widetilde{\mu}}\right\},
\end{align}
$\forall t\in[0,t_{\max})$, and by taking the inverse logarithm: 
\begin{align} \label{eq:bar ksi psi}
		0 \leq -\underline{\xi}_\psi \leq \xi_{\psi_k}(t) \leq \bar{\xi}_\psi < 1,
\end{align}
$\forall k\in\mathcal{K}$, where $\bar{\xi}_\psi \coloneqq \tfrac{\exp(\bar{\varepsilon}_\psi)-1}{\exp(\bar{\varepsilon}_\psi)}$ and $\underline{\xi}_\psi \coloneqq \tfrac{\exp(-\bar{\varepsilon}_\psi)-1}{\exp(-\bar{\varepsilon}_\psi)}$. Note that $\bar{B}_\psi$ as well as $\varepsilon_\psi(0)$ are finite, due to the choice $\psi_k(0) < \rho_{\psi_k}(0) < 2$, $\forall k\in\mathcal{K}$. Hence, since $\widetilde{\mu}$ is strictly positive, $\bar{\varepsilon}_\psi$ is also finite. Therefore, we conclude the boundedness of $\varepsilon_{\psi_k}, r_{\psi_k}(\xi_{\psi_k}(t))$, $e_v(t)$ in compact sets, $\forall k\in \mathcal{K}$, and therefore, through \eqref{eq:vel_i_des}, the boundedness of $\omega_{i,\text{des}}(t)$, $\forall i\in\mathcal{N}, t\in[0,t_{\max})$. From the proven boundedness of $p(t)$ and $p_{i,\text{des}}(t)$, we also conclude the boundedness of $n(x(t),t)$ and invoking $\widetilde{v} = v + n(x,t) = e_v(t)-v_{\text{des}}(t)$ and \eqref{eq:ksi v tau_max}, the boundedness of $v(t)$ and $\dot{x}(t)$,  $\forall t\in[0,t_{\max})$. 
Moreover, in view of \eqref{eq:bar epsilon e}, \eqref{eq:bar ksi e}, \eqref{eq:closed loop vector}, \eqref{eq:vel_i_des}, we also conclude the boundedness of $\dot{v}_{\text{des}}(t)$. 

Proceeding along similar lines, we consider the positive definite Lyapunov candidate $V_v:\mathbb{R}\to\mathbb{R}_{\geq 0}$ with $V_v(\varepsilon_v) = \tfrac{1}{2}\varepsilon^\top_v\Gamma\varepsilon_v$. By computing $\dot{V}_v(\varepsilon_v) = \left[ \frac{\partial V_v(\varepsilon_v)}{\partial \varepsilon_v}\right]^\top\dot{\varepsilon}_v$ and using the dynamics $\dot{\xi}_v = \rho_v(t)^{-1}(\dot{e}_v(t)$ $- \dot{\rho}_v(t)\xi_v)$, we obtain:

\begin{align}
&\dot{V}_v(\varepsilon_v) = - \varepsilon^\top_v \Sigma_v(\xi_v,t)\Gamma M^{-1}\Gamma\Sigma_v(\xi_v,t)\varepsilon_v \notag\\
& -\varepsilon^\top_v \Sigma_v(\xi_v,t) \Big\{\Gamma M^{-1}\Big[C(v)v + g(x) + w(x,v,t) \Big] \notag\\
&\hspace{25mm}- \dot{n}(x,t) + \dot{v}_{\text{des}} + \dot{\rho}_v(t)\xi_v \Big\}.  \label{eq:V_v 1}
\end{align}
Since we have proven the boundedness of $v(t)$ and $\dot{x}$, $\forall t\in[0,t_{\max})$ the terms  $C(v)v$, $\dot{n}(x,t)$, and $w(x,v,t)$ are also bounded, $t\in[0,t_{\max})$, due to the continuities of $C(\cdot)$, $w(\cdot)$, and $\dot{n}(\cdot)$ in $v$, $\dot{x}$ and the boundedness of $w(\cdot)$ and $\dot{n}()$ in $x,t$. Moreover, $g(x)$, $\xi_v(t)$, and $\dot{\rho}_{v}(t)$ are also bounded due to \eqref{eq:M property 2}, \eqref{eq:ksi v tau_max}, and by construction, respectively. By also using \eqref{eq:M property 1}, we obtain from \eqref{eq:V_v 1}: $\dot{V}_v(\varepsilon_v) \leq  - \underline{\lambda}_K \|\Sigma_v(\xi_v,t) \varepsilon_v \|^2 + \|\Sigma_v(\xi_v,t) \varepsilon_v \| \bar{B}_v$, where $\bar{B}_v$ is a positive term, independent of $t_{\max}$, satisfying $\bar{B}_v \geq \Big\| \frac{\max_{i\in\mathcal{N}}\{\gamma_i\}}{\min_{i\in\mathcal{N}}\{\underline{m}_i\}} \Big[C(v)v + g(x) + w(x,v,t) \Big] -\dot{n}(x,t) + \dot{v}_{\text{des}}(t) + \dot{\rho}_v(t)\xi_v(t) \Big\|$, $\forall t\in[0,t_{\max})$ and $\underline{\lambda}_K \coloneqq \frac{\min_{i\in\mathcal{N}}\{\gamma_i\}^2}{\max_{i\in\mathcal{N}}\{\bar{m}_i\}} > 0$.
Hence, $\dot{V}_v(\varepsilon_v) < 0 \Leftrightarrow \|\Sigma_v(\xi_v,t) \varepsilon_v \| > \tfrac{\bar{B}_v}{\underline{\lambda}_{K}}$. By noting that $r_{v}(x) = \frac{\partial T_{v}(x)}{\partial x} = \frac{2}{(1+x)(1-x)} > 2 >1$,
$\forall x\in(-1,1)$, as well as $\rho_{v_{i,\ell}}(t) \leq \rho_{v^0_{i,\ell}}$, $\forall \ell\in\{1,\dots,6\},t\in\mathbb{R}_{\geq 0}$, we conclude that $\|\Sigma_v(\xi_v(t),t)\varepsilon_v(t) \|$ $=$ $\sqrt{\sum_{i\in\mathcal{N}} \sum_{\ell\in\{1,\dots,6\}}\frac{[r_v(\xi_{v_{i,\ell}}(t))]^2}{[\rho_{v_{i,\ell}}(t)]^2} [\varepsilon_{v_{i,\ell}}(t)]^2} \geq \frac{1}{\widetilde{\rho}}\|\varepsilon_v(t)\|$, $\forall t\in[0,t_{\max})$, where $\widetilde{\rho} \coloneqq \max\limits_{\stackrel{i\in\mathcal{N}}{m\in\{1,\dots,6\}} }\{ \rho_{\scr v^0_{i,m}} \}$. 
Hence, we conclude that $\dot{V}_v(\varepsilon_v) < 0, \forall \| \varepsilon_v \| \geq  \frac{\widetilde{\rho} \bar{B}_v}{\underline{\lambda}_K},  \forall t\in[0,t_{\max})$,
 and consequently that 
 
\begin{align} \label{eq:bar epsilon v}
	\| \varepsilon_v(t)\| \leq \bar{\varepsilon}_v \coloneqq \max\left\{ \varepsilon_v(0), \frac{\widetilde{\rho} \bar{B}_v}{\underline{\lambda}_K} \frac{\max\limits_{i\in\mathcal{N}}\{\gamma_i\}}{\min\limits_{i\in\mathcal{N}}\{\gamma_i\}} \right\},
\end{align}
$\forall t\in[0,t_{\max})$,
and by taking the inverse logarithm function: 
\begin{equation} \label{eq:bar ksi v}
	-1 < -\bar{\xi}_v \leq \xi_{v_{i, \ell}}(t) \leq \bar{\xi}_v < 1,
\end{equation}
$\forall \ell \in\{1,\dots,6\}$, $t\in[0,t_{\max})$ where $\bar{\xi}_v \coloneqq \tfrac{\exp(\varepsilon_v)-1}{\exp(\varepsilon_v)+1} = -\tfrac{\exp(-\varepsilon_v)-1}{\exp(-\varepsilon_v)+1}$. Note that the terms $\bar{B}_v$ finite, $\forall t\in[0,t_{\max})$. Moreover, the term $\varepsilon_v(0)$ is finite due to the choice $\rho_{\scr v^0_{i,\ell}} > |e_{v_{i,\ell}}(0)|, \forall \ell \in\{1,\dots,6\},i\in\mathcal{N}$. Hence, since $\underline{\lambda}_K$ is strictly positive, the term $\bar{\varepsilon}_v$ is also finite.
Thus,  the terms $e_v(t)$, $\widetilde{r}_v(\xi_v(t)) $ and hence the control laws \eqref{eq:u_i} are also bounded in compact sets for all $t\in[0,t_{\max})$. What remains to be shown is that $t_{\max} = \infty$. Towards that end, suppose that $t_{\max}$ is finite, i.e., $t_{\max} < \infty$. Then, according to Theorem \ref{thm:forward_completeness}, it holds that 
\begin{equation}
	L \coloneqq	\lim\limits_{t \to  t^{-}_{\max}}\Big[\| z(t) \| + \frac{1}{d_\mathcal{S}((z(t),t),\partial \Omega)} \Big] = \infty, \label{eq:L t_max < infty}
\end{equation}
where $\|z\|\coloneqq \|p\| + \|v\| + \|R\|_T$ and, with a slight abuse of notation with respect to Section \ref{sec:preliminaries}, $d_\mathcal{S}((z(t),t),\partial \Omega) \coloneqq \inf_{(z'_{p,v},R')\in\partial \Omega} \{ \|z_{p,v} - z'_{p,v}\| + \|R - R'\|_T \}$, and $z_{p,v} \coloneqq [p^\top,v^\top]^\top\in\mathbb{R}^{3N}\times\mathbb{R}^{6N}$. We now aim to prove that \eqref{eq:L t_max < infty} is a contradiction. Firstly, it holds that $\|R(t)\|_T = \sum_{i\in\mathcal{N}}\|R_i(t)\|_F \leq N \sup_{t\in [0,t_{\max})}\{ \max_{i\in\mathcal{N}}\{R_i(t)\}\}$. However, according to Proposition \ref{prop:R trace}, it holds that $-1 \leq \text{tr}(R) \leq 3$ for any $R\in\mathbb{SO}(3)$. Hence, $\|R(t)\|_T \leq 3N, \forall t\in[0,t_{\max}]$. Moreover, from \eqref{eq:bar ksi v} and \eqref{eq:ksi_i_v} we obtain $\|e_v(t)\| \leq \sqrt{6}\bar{\xi}_v\widetilde{\rho}$, $\forall t\in[0,t_{\max})$. By invoking \eqref{eq:bar epsilon e}, \eqref{eq:bar epsilon psi}, we can also conclude that there exists a finite $\bar{v}_{\text{des}}$ such that $\|v_{\text{des}}(t)\| \leq \bar{v}_{\text{des}}$, $\forall t\in[0,t_{\max})$. Therefore, since $\|n_i(x_i,t)\| \leq \bar{n}_i$, $\forall x_i\in\mathbb{SE}(3),t\in\mathbb{R}_{\geq 0}, i\in\mathcal{N}$, $v = \widetilde{v} - n(x,t) = e_v + v_{\text{des}} -n(x,t)$ implies that there exists a finite $\bar{v}$ such that $\|v(t)\| \leq \bar{v}$, $\forall t\in[0,t_{\max})$. Hence, $\| p(t)\| = \| \int_{0}^{t_{\max}}\bar{R}(s)v(s)ds\| \leq \int_{0}^{t_{\max}}\|\bar{R}(s)v(s)\|ds = \int_{0}^{t_{\max}}\|v(s)\|ds \leq  \int_{0}^{t_{\max}}\bar{v}ds \Rightarrow \|p(t)\| \leq t_{\max}\bar{v}$, $\forall t\in[0,t_{\max})$, which proves the boundedness of $\|p(t)\|$, since $t_{\max}$ is bounded. Next, note that $\partial \Omega = \{(p,v,R,t)\in \mathbb{R}^{3N}\times\mathbb{R}^{6N}\times\mathbb{SO}(3)^N\times\mathbb{R}_{\geq 0} : ( \exists k\in\mathcal{K} : \xi_{e_k}(p_{k_1},p_{k_2},t) = -C_{k,\text{col}}$ or $\xi_{e_k}(p_{k_1},p_{k_2},t) = C_{k,\text{con}}$ or $\xi_{\psi_k}(R_{k_1},R_{k_2},t) = 1)$ or $ (\exists i\in\mathcal{N},\ell\in\{1,\dots,6\} : \xi_{v_{i,\ell}}(x,v_i,t) = -1 \text{ or }  \xi_{v_{i,\ell}}(x,v_i,t) = 1 ) \}$. We have proved, however, from \eqref{eq:bar ksi e}, \eqref{eq:bar ksi psi}, and \eqref{eq:bar ksi v} that the maximal solution satisfies the strict inequalities $-C_{k,\text{col}} < -\underline{\xi}_e \leq \xi_{e_k}(p_{k_1}(t),p_{k_2}(t),t) \leq \bar{\xi}_e < C_{k,\text{con}}$, $\xi_{\psi_k}(R_{k_1}(t),R_{k_2}(t),t) \leq \bar{\xi}_{\psi} < 1$, and $|\xi_{v_{i,\ell}}(x(t),v_i(t),t)| \leq \bar{\xi}_v < 1 $, $\forall k\in\mathcal{K}$, $\ell\in\{1,\dots,6\}$, $i\in\mathcal{N}$, $t\in[0,t_{\max})$. Therefore, we conclude that there exists a strictly positive constant $\epsilon_z$, $\in\mathbb{R}_{>0}$ such that $d_\mathcal{S}((z(t),t),\partial \Omega) \geq \epsilon_z$. Therefore, we have proved that $L \leq (t_{\max} + 1)\bar{v} + 3N + \epsilon_z^{-1}$, which is finite, since $t_{\max}$ is finite. This contradicts \eqref{eq:L t_max < infty} and hence, we conclude that $t_{\max} = \infty$.

We have proved the containment of the errors $e_k(t)$, $\psi_k(t)$ in the domain defined by the prescribed performance funnels: $-C_{k,\text{col}}\rho_{e_k}(t) < e_k(t) < C_{k,\text{con}}\rho_{e_k}(t)$, $0 \leq \psi_k(t) < \rho_{\psi_k}(t)$,
$\forall k\in \mathcal{K}$, $t\in \in\mathbb{R}_{\geq 0}$, which also implies that: $d_{k,\text{col}} < \| p_{k_1}(t) - p_{k_2}(t) \| < d_{k,\text{con}}$, $0 \leq \psi_k(t) < 2$, $\forall k\in \mathcal{K}$, $t\in\mathbb{R}_{\geq 0}$, i.e., avoidance of the singularity $\psi_k = 2$ and satisfaction of the distance and connectivity constraints for the initially connected edge set $\mathcal{E}$. The closed loop signals and functions are also proven to be bounded for all $t\in[0,\infty)$, which leads to the conclusion of the proof.
\end{pf}

\begin{remark} [\textbf{Prescribed performance}]
	We can deduce from the aforementioned proof that the proposed control scheme achieves its goals without resorting to the need of rendering $\bar{\varepsilon}_e$, $\bar{\varepsilon}_\psi$, $\bar{\varepsilon}_v$ arbitrarily small by adopting extreme values of the control gains $\delta_i, \gamma_i$. Notice that \eqref{eq:bar epsilon e}, \eqref{eq:bar epsilon psi}, and \eqref{eq:bar epsilon v} hold no matter how large the finite bounds $\bar{\varepsilon}_e$, $\bar{\varepsilon}_\psi$, $\bar{\varepsilon}_v$ are. Hence, the actual performance of the system is determined solely by the performance functions $\rho_{e}(t), \rho_{\psi}(t), \rho_v(t)$ and the parameters $C_{k,\text{col}}, C_{k,\text{con}}$, as mentioned in Remark \ref{rem:pp functions and gain tuning}.
\end{remark}

%
%
%

\begin{figure*}
	\centering
	\includegraphics[width=\textwidth]{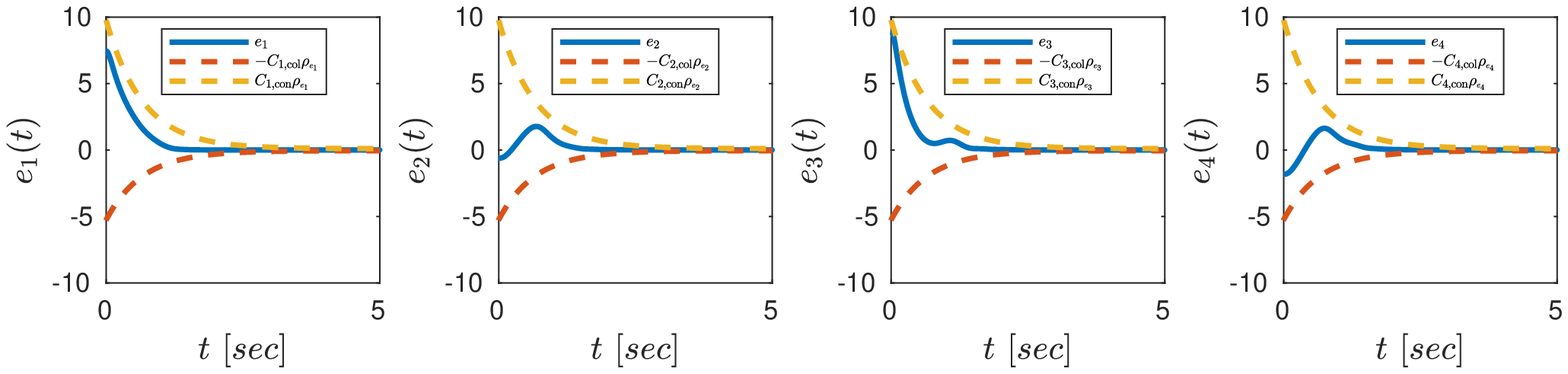}
	\caption{The distance errors $e_k(t)$ along with the performance functions $-C_{k,\text{col}}\rho_{e_k}(t)$, $C_{k,\text{con}}\rho_{e_k}(t)$, $\forall k\in\mathcal{K}$.}
	\label{fig:dist_errors}
\end{figure*}

\begin{figure*}
	\centering
	\includegraphics[width=\textwidth]{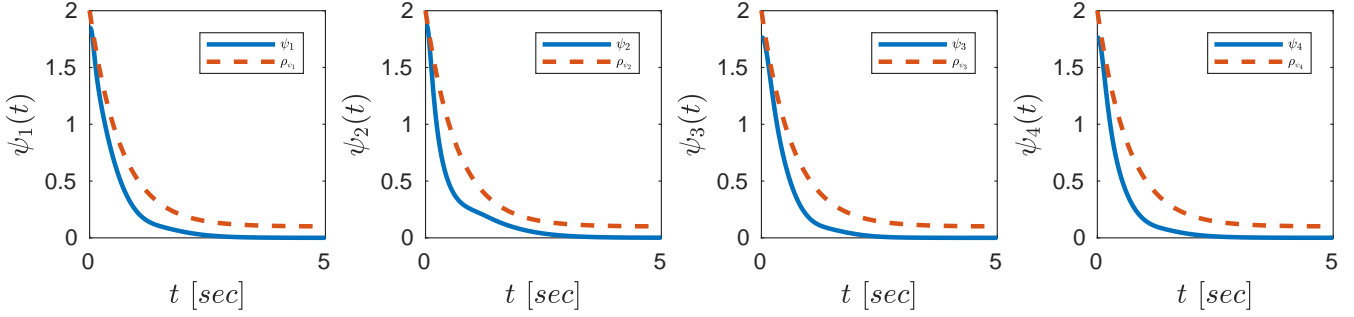}
	\caption{The orientation errors $\psi_k(t)$ along with the performance function $\rho_{\psi_k}(t)$, $\forall k\in\mathcal{K}$.}
	\label{fig:orient_errors}
\end{figure*}

\begin{figure*}
	\centering
	\includegraphics[width=\textwidth]{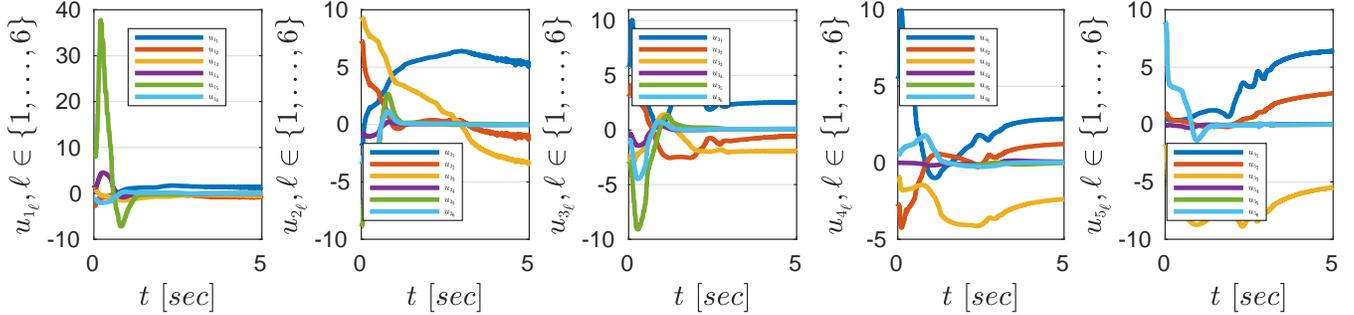}
	\caption{The control inputs of the agents $u_i(t)$, $\forall i\in\mathcal{N}$.}
	\label{fig:inputs}
\end{figure*}


\section{Simulation Results} \label{sec:simulation_results}

We considered $N=5$ spherical agents with $\mathcal{N} = \{1,\dots,5\}$, with 
dynamics of the form \eqref{eq:system}, with $r_i=1 \text{m}$, $s_i = 4 \text{m}$, and dynamic parameters (mass and moment of inertia) randomly selected in $(0,1)$, $i\in\mathcal{N}$. We selected the exogenous disturbances and measurement noise as $w_i = A_{w_i}\sin(\|p_1\|_1  \text{tr}(R_i)\omega_{w,i}t + \phi_{w,i})v_i$, and $n_i = A_{n_i}\sin(\|p_1\|_1  \text{tr}(R_i)\omega_{n,i}t + \phi_{n,i})v_i$, where the parameters $A_{w_i}, A_{n_i}, \omega_{w,i}, \omega_{n,i}, \phi_{w,i}, \phi_{n,i}$ are randomly chosen in $(0,0.1)$, $\forall i\in\mathcal{N}$. The initial conditions were taken as: $p_1(0)=[0,0,0]^\top \ \text{m}$, $p_2(0)=[-2.1,-2.3,2]^\top \ \text{m}$, $p_3(0)=[1.3,1.3,1.5]^\top \ \text{m}$, $p_4(0)=[-2,3.25,2.2]^\top \ \text{m}$, $p_5(0)=[2,2.4,-0.15]^\top \ \text{m}$, $R_1(0)= R_4(0) = R_5(0) = I_{3}$, $R_2(0)$ $=$ $[-0.8253$,$0$,$0.5646$;$0$,$1$,$0.2562$;$-0.5646$,$0$,$-0.8253]$,
$R_3(0)$ $=$ $[-0.3624$,$0$,$0.9320$;$0.6591$,$0.7071$,$0.2562$;$-0.6591$, $0.7071$,$-0.2562]$,
$v_1(0) = v_2(0 = v_3(0) = v(4) = 0_{6 \times 1}$, which form the edge set $\mathcal{E} = \{\{1,2\}$, $\{1,3\}$, $\{3,4\}$, $\{3,5\} \}$. The desired graph formation was defined by the constants $d_{k,\text{des}} = 2.5\text{m}$ and $R_{k,\text{des}} = [0.5,-0.8660, 0;0.6124, 0.3536, -0.7071; 0.6124,0.3536$, $0.7071]$,
$\forall k\in\{1,\dots,4\}$. We selected $d_{k, \text{col}} = 2$, $d_{k, \text{con}} = 4$, and in view of \eqref{eq:C_k}, $C_{k, \text{col}} = 2.25$ and $C_{k, \text{con}} = 9.75$. Moreover, the parameters of the performance functions were chosen as $\rho_{\scr e_k,\infty} $ $= \rho_{\scr \psi_k,\infty} = 0.1$, $\rho_{\scr \psi_k,0} = 1.99 > \max\{\rho_{\scr \psi_1}(0), \rho_{\scr \psi_2}(0), \rho_{\scr \psi_3}(0)\}$ and $l_{e_k} = l_{\psi_k} = 1.5$. In addition, we chose $\rho_{\scriptscriptstyle v^{0}_{i,\ell}} = 2 | e_{v_{i,\ell}}(0)| + 1$, $l_{v_{i,\ell}} = 0.2$ and $\rho_{\scriptscriptstyle v^{\infty}_{i,\ell}} = 0.1$, for every $i \in \mathcal{N}$, $\ell \in \{1, \dots, 6\}$. Finally, the control gains were set to $\delta_i = 0.1$ and $\gamma_i = 15$, $\forall i\in\mathcal{N}$.
The simulation results are shown in Figs. \ref{fig:dist_errors}-\ref{fig:inputs}.
In particular, Figs. \ref{fig:dist_errors} and \ref{fig:orient_errors} depict the distance and orientation errors $e_k(t)$, $\psi_k(t)$, respectively, along with the corresponding performance functions $\rho_k(t)$, $\rho_{\psi_k}(t)$, $\forall k\in\mathcal{K}$. 
Moreover, Fig. \ref{fig:inputs} depict the control inputs of the agents, $\forall t\in [0,5]$ seconds. 
It can be observed that, although the initial errors $e_{k}(0)$ and $\psi_k(0)$ are very close to the performance bounds, the proposed control algorithm achieves convergence to the desired formation configuration in a short time interval without significant control effort. A video illustrating the simulation results can be found in \textit{https://www.youtube.com/watch?v=Z4xLyO1twvk}. 

\section{Conclusions and Future Work} \label{sec:conclusions}

In this paper we proposed a robust decentralized control protocol for distance- and orientation-based formation control of multiple rigid bodies with unknown dynamics in the special Euclidean group $\mathbb{SE}(3)$. The proposed control protocol guarantees collision avoidance and connectivity maintenance with the initially connected agents. Moreover, the transient- and steady-state trajectories of the closed loop system are determined by pre-specified performance functions. Simulation examples have verified the efficiency of the proposed approach. Future efforts will be devoted towards extending the current results to collision avoidance among all the agents as well as collision avoidance with obstacles in the environment. 

%

\bibliographystyle{unsrt}        
\bibliography{references}           

\appendix

\end{document}